\documentclass[a4paper,fleqn]{cas-dc}
\usepackage[numbers,sort&compress]{natbib}
\def\tsc#1{\csdef{#1}{\textsc{\lowercase{#1}}\xspace}}
\tsc{WGM}
\tsc{QE}
\tsc{EP}
\tsc{PMS}
\tsc{BEC}
\tsc{DE}

\usepackage{optidef}

\usepackage{hyperref}
\usepackage{algorithm}
\usepackage{algpseudocode}
\usepackage{amsmath, amssymb, amsfonts, amsthm}
\usepackage{graphicx}
\usepackage{color}
\usepackage{bm}
\usepackage{float}
\usepackage{multirow}
\usepackage{setspace}
\usepackage{textcomp}
\usepackage{times}
\usepackage[section]{placeins}
\usepackage{rotating}
\usepackage{hhline}
\usepackage{subfig}
\usepackage{booktabs}
\usepackage{ragged2e}
\usepackage{epstopdf}

\usepackage{mathtools}
\usepackage{longtable}

\graphicspath{{./figs/}}
\usepackage[labelfont=bf,labelsep=colon]{caption}

\begin{document}
	\let\WriteBookmarks\relax
	\def\floatpagepagefraction{1}
	\def\textpagefraction{.001}
	\graphicspath{{./figs/}}
	\shortauthors{A. Abubakar, S. E. Assabil, A. Hussain, V.-H. Bui}
	
	\title[mode=title]{\bf
		Decision-Gated Surrogate-Assisted Stochastic Optimization with Independent High-Fidelity Certification for Photovoltaic Hosting-Capacity Planning
	}
	
	% ============================================================
	% AUTHORS
	% ============================================================
	
	\author[1,2]{Ali Abubakar}[
	orcid=0000-0002-8727-3487]
	\cormark[2]
	\ead{aliabu@umich.edu}
	
	\author[5]{Samuel Essamuah Assabil}[
	style=english,
	orcid=0000-0003-3891-8530]
	\ead{samuel.assabil@ucc.edu.gh}
	
	\author[4]{Akhtar Hussain}
	\ead{akhtar.hussain@ulaval.ca}
	
	\author[1,3]{Van-Hai Bui}
	\cormark[1]
	\ead{vhbui@umich.edu}
	
	% ============================================================
	% AFFILIATIONS
	% ============================================================
	
	\affiliation[1]{
		organization={Michigan Institute for Data and AI in Society (MIDAS),
			University of Michigan},
		city={Ann Arbor},
		state={Michigan},
		country={USA}
	}
	
	\affiliation[2]{
		organization={Department of Mathematics,
			University of Cape Coast},
		city={Cape Coast},
		country={Ghana}
	}
	
	\affiliation[3]{
		organization={Department of Electrical and Computer Engineering,
			College of Engineering and Computer Science,
			University of Michigan--Dearborn},
		city={Dearborn},
		state={Michigan},
		country={USA}
	}
	
	\affiliation[4]{
		organization={Department of Electrical and Computer Engineering,
			Université Laval},
		city={Québec City},
		state={Québec},
		country={Canada}
	}
	
	\affiliation[5]{
		organization={Department of Statistics,
			University of Cape Coast},
		city={Cape Coast},
		country={Ghana}
	}
	
	% ============================================================
	% CORRESPONDING AUTHOR
	% ============================================================
	
	\cortext[cor1]{\textbf{Corresponding author: Van-Hai Bui
			(vhbui@umich.edu)}}

\begin{abstract}
High photovoltaic (PV) penetration planning requires repeated nonlinear PV and power-flow evaluations under uncertain irradiance and demand, particularly when two competing objectives, penetration and total electrical loss, are optimized simultaneously subject to reliability constraints. This paper develops a decision-focused, surrogate-assisted stochastic multiobjective framework in which a calibrated two-diode PV model coupled with a distribution-network simulator provides the high-fidelity reference, while matched data-only and physics-informed surrogates accelerate optimization and are evaluated empirically for decision reliability across the loss–penetration trade-off. Although the two surrogates exhibit nearly identical held-out errors, optimizer-realistic screening yields markedly different operational validity rates of 44.4\% and 93.4\%, respectively, motivating selection of the physics-informed model on decision performance rather than test accuracy. Surrogate inference cuts the cost of a 60-scenario design evaluation from 286.0 to 0.46~ms, an approximately $620\times$ speedup. A decision-gated enrichment strategy further targets Pareto-critical, constraint-boundary, and surrogate-disagreement regions, reducing prediction error by 39.1\%. Independent 500-scenario certification verifies all 80 screened candidates and identifies 46 non-dominated designs; the selected compromise attains 70.74\% expected penetration with 297.9~kW expected total electrical loss. Cross-feeder transfer eliminates baseline voltage violations on the 85- and 69-bus systems, but produces overvoltage and a 62.7\% increase in feeder-network loss on the 33-bus system, demonstrating that certified penetration is topology-specific and demands decision-level auditing, independent certification, and feeder-specific validation.
\end{abstract}

	
	\maketitle
	
\begin{itemize}
	\item Chance-constrained multi-objective search balances penetration and system losses.
	\item Physics-informed surrogate selection relies on decision-level high-fidelity audits.
	\item Periodic high-fidelity enrichment rejects updates that worsen decisions.
	\item Independent 500-scenario certification precedes compromise design selection.
	\item Cross-feeder transfer reveals topology-dependent voltage and loss limits.
\end{itemize}

	\section{Background}
	\label{sec:intro}	
\noindent	High photovoltaic (PV) penetration reshapes the operating regime of a
	distribution feeder: voltage profiles shift, power flow can reverse toward
	the substation, thermal loading on lines and transformers can change
	substantially, and system losses respond nonlinearly to interactions among
	irradiance, temperature, and demand \cite{lyetal2025chanceHC,Kaseb2026,PICNN2024}.
	Because irradiance and demand are stochastic, these effects cannot be characterized at a single deterministic operating point; they must be evaluated across a distribution of plausible weather and demand scenarios \cite{Kaseb2026}. Prior work has responded with scenario-based simulation, probabilistic and possibilistic uncertainty representations, distributionally robust and chance-constrained formulations, sample-average approximation, and multi-objective search \cite{deakin2019stochastic,arshad2019hostingcapacity,Wang2025,Che2025,ulhassan2022benchmark,Herding2024,alietal2022possibilistic,lyetal2025chanceHC,shen2025klhc,zhang2023wasserstein,wang2025dlhosting}, with reduced-order representations such as polynomial chaos expansion controlling the cost of repeated uncertainty propagation elsewhere \cite{claeys2022gpc}. Across this literature, one difficulty is structural: once uncertainty is represented explicitly, every candidate a stochastic optimizer considers must be evaluated over many scenarios, and a full two-diode PV representation coupled to a nonlinear AC power flow is expensive at that scale \cite{Xu2026}. Reducing physical fidelity for computational speed can compromise the voltage, loading, loss, and hosting-capacity estimates that govern design acceptance \cite{Zhang2025Simplifications,Seidaliseifabad2019HC,Ma2025HC}.\\
	
\noindent	Neural surrogates provide a natural means of reducing the computational burden of repeated high-fidelity evaluations, while physics-informed variants further embed governing-equation residuals into the training objective and have shown promise in power-system state estimation, optimal power flow, and dynamic simulation \cite{raissi2019pinn,cuomo2022pinnreview,huangwang2023pinnreview,Xu2026,misyris2020pinn,nellikkath2022pinn,Kaseb2026,Nadal2025,PICNN2024,stiasny2024pinnsim}. However, incorporating a physical residual does not guarantee superior surrogate performance: its effectiveness depends on the available data, model architecture, residual formulation and weighting, and the region of the input--output space being approximated. This limitation becomes particularly important in optimization, where surrogate evaluations are not distributed like the training samples but progressively concentrate around Pareto-critical regions and active constraint boundaries \cite{Xu2026}. Consequently, errors that appear negligible under aggregate test metrics can become decision-critical. A small error in minimum voltage, for example, may have little consequence for a clearly feasible design yet reverse the feasibility classification of a candidate near an operational limit; similarly, modest objective errors can alter dominance relationships and reorder solutions along a Pareto front.\\
	
\noindent	Predictive fidelity and optimization reliability should therefore be treated as distinct properties of a surrogate. Models with comparable held-out errors may exhibit substantially different behavior in the regions that determine feasibility and trade-off decisions. Surrogate assessment should accordingly extend beyond average prediction error to include decision-relevant measures such as feasibility agreement, false-feasibility rate, and preservation of Pareto structure under high-fidelity re-evaluation \cite{stiasny2024pinnsim}. This consideration is especially important when comparing purely data-driven and physics-informed surrogates: physical regularization may promote consistency with selected governing relationships, but it does not by itself establish superior decision performance. A rigorous comparison consequently requires models matched in training data, architecture, and computational budget, together with evaluation in optimizer-relevant regions where prediction errors have the greatest engineering consequence \cite{lyetal2025chanceHC}.\\
	
\noindent A complementary challenge in surrogate-assisted power-system optimization is
how to combine computational efficiency with the physical authority of the
underlying network model. Physics-informed learning can improve consistency
with governing power-system relationships and reduce constraint violations,
as demonstrated in surrogate formulations for AC optimal power flow
\cite{nellikkath2022pinn}, while physics-informed simulation has also been
used to accelerate computationally intensive power-system dynamics
\cite{stiasny2024pinnsim}. Related work in stochastic PV hosting-capacity
assessment has similarly used reduced-order or learned models to avoid
repeated power-flow solutions under uncertainty
\cite{claeys2022gpc,lyetal2025chanceHC}. These developments support a
hierarchical view of surrogate modeling: low-cost models are valuable for
exploration, but the high-fidelity simulator remains the appropriate reference
for physically consequential feasibility and reliability decisions. This distinction becomes particularly important because surrogate error is
generally nonuniform over the design space. In surrogate-based and
multi-fidelity optimization, adaptive sampling is therefore commonly used to
return informative or poorly resolved designs to a higher-fidelity model,
augment the available training information, and progressively improve the
approximation in regions relevant to optimization \cite{nellikkath2022pinn}. Such strategies typically
alternate surrogate fitting, acquisition of informative samples, and
high-fidelity evaluation rather than relying on a fixed approximation
throughout the search. More generally, multi-fidelity optimization has evolved
around this division of roles between inexpensive approximations and costly
but authoritative evaluations, with fidelity management determining when
higher-fidelity information is sufficiently valuable to justify its cost.
For constrained stochastic power-system planning, however, improved average
prediction accuracy is not by itself sufficient evidence that a surrogate
update is beneficial \cite{wang2025dlhosting,shen2025klhc}. An update may reduce aggregate error while worsening
feasibility classification or altering dominance relationships near the
Pareto front. \\

\noindent
The literature most closely related to this study spans three complementary
strands. The first addresses stochastic PV hosting-capacity and penetration
planning, where uncertainty in irradiance, demand, and operating conditions is
represented through probabilistic, possibilistic, distributionally robust, or
sample-based formulations. The second concerns data-driven and physics-informed
surrogates for power-system state estimation, optimal power flow, and dynamic
simulation, motivated by the computational cost of repeated nonlinear network
solutions. The third comprises surrogate-assisted and multi-fidelity optimization
methods that return informative samples to a high-fidelity model to improve
approximation quality in search-relevant regions. Table~\ref{tab:litmatrix}
summarizes representative studies across seven attributes characterizing this
intersection: two-diode PV modeling, AC power-flow representation, stochastic
uncertainty, multiobjective optimization, surrogate acceleration,
physics-informed learning, and high-fidelity enrichment. Among the studies
reviewed, no single formulation combines all seven within a unified stochastic
multiobjective PV planning framework. Earlier work by the authors established
physical, stochastic, and multiobjective foundations
\cite{abubakar2023pvpenetration,abubakar2024stochastic,abubakar2024harmonic},
but did not incorporate surrogate acceleration, physics-informed learning, or
decision-gated high-fidelity enrichment. Conversely, physics-informed
power-system studies have primarily focused on state prediction, optimal power
flow, and dynamic simulation
\cite{misyris2020pinn,nellikkath2022pinn,Kaseb2026,Nadal2025,PICNN2024,stiasny2024pinnsim},
while stochastic hosting-capacity and PV-impact studies have emphasized
uncertainty propagation, operational constraints, and feasibility limits
without combining decision-audited surrogate adaptation with independent
high-fidelity certification
\cite{Wang2025,Che2025,zhang2023wasserstein,wang2025dlhosting,lyetal2025chanceHC,hu2026tdcoord}.

\noindent
The resulting gap is therefore one of integration rather than any single
missing component. A decision-reliable framework must accelerate stochastic
multiobjective search without relinquishing the high-fidelity physical model
as the reference for feasibility, loss, and penetration; evaluate surrogate
quality in optimizer-relevant Pareto and constraint-boundary regions rather
than by average prediction error alone; and use targeted high-fidelity
enrichment with explicit safeguards on false feasibility and decision
agreement. Final reliability claims should consequently remain anchored to
independent high-fidelity evaluation. This motivates the architecture developed
here: vetted surrogate acceleration, periodic decision-aware high-fidelity
enrichment, and independent reliability verification within a unified
stochastic PV penetration--loss planning framework.
	\begin{table*}
		\centering
		\caption{Comparison of related studies in stochastic PV planning and
			physics-informed power-system surrogate modeling.
			\checkmark: explicitly incorporated; --: absent, unreported, or outside scope.}
		\label{tab:litmatrix}
		
		\vspace{-2pt}
		\scriptsize
		\setlength{\tabcolsep}{2.8pt}
		\renewcommand{\arraystretch}{0.98}
		
		\resizebox{0.99\textwidth}{!}{%
			\begin{tabular}{lccccccc}
				\toprule
				\textbf{Study} &
				\textbf{Two-diode PV} &
				\textbf{AC-PF} &
				\textbf{Stochastic} &
				\textbf{Multiobjective} &
				\textbf{Surrogate} &
				\textbf{Physics-informed} &
				\textbf{HF enrichment} \\
				\midrule
				
	Zhang et al.\ 2023 \cite{zhang2023wasserstein}
& -- & \checkmark & \checkmark & -- & -- & -- & -- \\

Claeys et al.\ 2022 \cite{claeys2022gpc}
& -- & \checkmark & \checkmark & -- & \checkmark\,(PCE) & -- & -- \\

Mulenga \& Etherden 2022 \cite{ulhassan2022benchmark}
& -- & \checkmark & \checkmark & -- & -- & -- & -- \\

Ali et al.\ 2022 \cite{alietal2022possibilistic}
& -- & \checkmark & \checkmark\,(possibilistic) & -- & -- & -- & -- \\

Nellikkath \& Chatzivasileiadis 2022 \cite{nellikkath2022pinn}
& -- & \checkmark & -- & -- & \checkmark & \checkmark & -- \\

Abubakar et al.\ 2023 \cite{abubakar2023pvpenetration}
& \checkmark & \checkmark & \checkmark & \checkmark
& -- & -- & -- \\

Stiasny et al.\ 2024 \cite{stiasny2024pinnsim}
& -- & \checkmark & -- & -- & \checkmark & \checkmark & -- \\

Herding et al.\ 2024 \cite{Herding2024}
& -- & \checkmark & \checkmark & -- & -- & -- & -- \\

Neural-network OPF embedding 2024 \cite{PICNN2024}
& -- & \checkmark & -- & -- & \checkmark & \checkmark & -- \\

Alfaris et al.\ 2024 \cite{new11}
& -- & \checkmark & -- & -- & -- & -- & -- \\

Bin Hudayb et al.\ 2024 \cite{new9}
& -- & \checkmark & -- & -- & -- & -- & -- \\

Jamahori et al.\ 2024 \cite{new10}
& -- & \checkmark & -- & -- & -- & -- & -- \\

Alfrd \& Ebshish 2024 \cite{new5}
& -- & \checkmark & -- & -- & -- & -- & -- \\

Abubakar et al.\ 2024 \cite{abubakar2024stochastic}
& \checkmark & \checkmark & \checkmark & \checkmark
& -- & -- & -- \\

Nadal et al.\ 2025 \cite{Nadal2025}
& -- & \checkmark & -- & -- & \checkmark & \checkmark & -- \\

Wang et al.\ 2025 \cite{Wang2025}
& -- & \checkmark & -- & -- & -- & -- & -- \\

Wang et al.\ 2025 \cite{wang2025dlhosting}
& -- & \checkmark & \checkmark & -- & \checkmark & -- & -- \\

Ly et al.\ 2025 \cite{lyetal2025chanceHC}
& -- & \checkmark & \checkmark & -- & \checkmark & -- & -- \\

Shen et al.\ 2025 \cite{shen2025klhc}
& -- & -- & \checkmark & -- & -- & -- & -- \\

Gulraiz et al.\ 2025 \cite{new2}
& -- & \checkmark & -- & -- & -- & -- & -- \\

Rajab et al.\ 2025 \cite{new6}
& -- & \checkmark & -- & -- & -- & -- & -- \\

Mohamed et al.\ 2025 \cite{new4}
& -- & \checkmark & -- & -- & -- & -- & -- \\

Che et al.\ 2025 \cite{Che2025}
& -- & -- & -- & -- & -- & -- & -- \\

Adak 2025 \cite{new8}
& -- & \checkmark & -- & -- & -- & -- & -- \\

Maghalseh et al.\ 2025 \cite{new15}
& -- & \checkmark & -- & -- & -- & -- & -- \\

Ali et al.\ 2025 \cite{new14}
& -- & \checkmark & -- & -- & -- & -- & -- \\

Qui\~{n}ones et al.\ 2025 \cite{new7}
& -- & -- & \checkmark & -- & -- & -- & -- \\

Kaseb et al.\ 2026 \cite{Kaseb2026}
& -- & \checkmark & -- & -- & \checkmark & \checkmark & -- \\

Shu et al.\ 2026 \cite{shu2026closedloop}
& -- & \checkmark & -- & -- & \checkmark & \checkmark & \checkmark \\

Hu et al.\ 2026 \cite{hu2026tdcoord}
& -- & \checkmark & -- & -- & -- & -- & -- \\

Abubakar et al.\ \cite{abubakar2024harmonic}
& -- & \checkmark & \checkmark & \checkmark
& -- & -- & -- \\
				
				\midrule
				\textbf{This work}
				& \checkmark
				& \checkmark
				& \checkmark
				& \checkmark
				& \checkmark
				& \checkmark
				& \checkmark \\
				\bottomrule
			\end{tabular}%
		}
		\vspace{-4pt}
	\end{table*}
	
	\noindent Against this background, the study makes three contributions that connect
	physical fidelity, computational efficiency, and statistical reliability
	within a single stochastic planning framework:
	
	\begin{enumerate}
		\item[I.]
		A chance-constrained stochastic \emph{multiobjective} formulation that
		couples a calibrated two-diode PV model with balanced AC power flow to
		jointly optimize expected PV penetration and expected total electrical
		loss under a prescribed reliability requirement. The high-fidelity model
		remains the authority for physical feasibility, loss, and penetration
		throughout the analysis.
		
		\item[II.]
		A decision-focused surrogate-selection and enrichment strategy in which
		matched data-only and physics-informed models are compared using both
		predictive accuracy and optimizer-realistic high-fidelity evidence.
		Periodic enrichment targets Pareto-critical, constraint-boundary, and
		surrogate-disagreement regions, while candidate updates are retained only
		when prediction-error, false-feasibility, and decision-agreement gates are
		simultaneously satisfied; otherwise, the previous surrogate is restored.
		
		\item[III.]
		Independent high-fidelity reliability certification of the screened
		candidate set using 500 scenarios and an exact one-sided 95\%
		Clopper--Pearson lower confidence bound, followed by compromise selection
		from the certified non-dominated set and cross-feeder stress testing.
		The transfer analysis explicitly examines whether a certified penetration
		level remains admissible under different feeder topologies rather than
		assuming that certification is directly portable across networks.
	\end{enumerate}
	
\noindent	\textbf{Terminology.}
	Throughout this paper, \emph{certification} denotes computational reliability
	certification. A design is certified only after independent high-fidelity
	evaluation confirms the prescribed planning limits and the one-sided 95\%
	Clopper--Pearson lower confidence bound on operational safety satisfies the
	required reliability threshold. The term does not refer to regulatory,
	equipment, or hardware certification. \vspace{-10pt}

	\section{Methodology}
	\label{sec:methodology}
	
\begin{figure}
		\centering
		\includegraphics[width=\columnwidth,height=8.4cm,keepaspectratio]{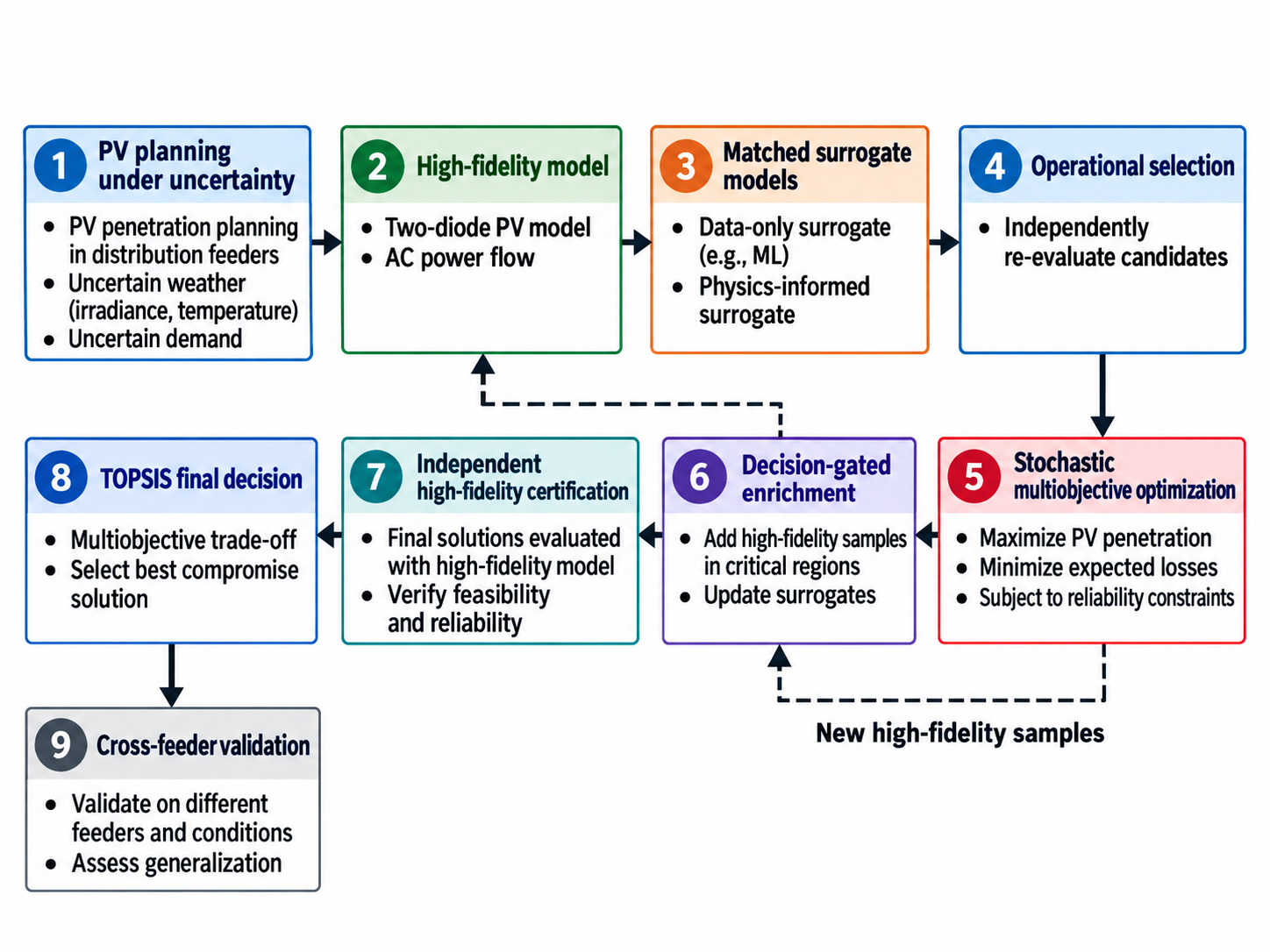}
		\caption{Decision-gated surrogate-assisted stochastic planning and
			certification workflow.}
		\label{fig:flow}
	\end{figure}

	\noindent
	The framework follows the evidence chain in Fig.~\ref{fig:flow}. During
	optimization, Pareto-critical, constraint-boundary, and surrogate-disagreement
	regions are periodically returned to the HF model. Surrogate updates are
	retained only when the prescribed decision-fidelity gates are satisfied;
	otherwise, the previous model is restored. The surrogate accelerates search,
	while HF simulation remains the reference for feasibility and certification.
	
	\subsection{High-Fidelity PV--Distribution-Network Model}
	\label{sec:physical_model}
	
	\noindent
	The HF model couples a calibrated two-diode PV representation with a balanced
	radial AC network. The 13-dimensional mixed-integer design vector is
	\begin{equation}
		\mathbf{x}=
		[N_p,N_s,S_c,I_c,\beta,\gamma,A_{\mathrm{dc}},A_{\mathrm{ac}},
		b_{\mathrm{PCC}},K,L,D,H]^T ,
		\label{eq:decision_vector}
	\end{equation}
\noindent	covering array topology, inverter sizing, orientation, conductor sizing,
	PCC location, and row geometry, subject to
	\begin{equation}
		\begin{aligned}
			\mathcal X=\{\mathbf{x}:~&
			(S_c,I_c)\in\mathcal C_{\mathrm{inv}},~
			N_p,N_s\in\mathbb Z_{>0},~
			b_{\mathrm{PCC}}\in\mathcal B,\\
			&A_{\mathrm{dc}},A_{\mathrm{ac}}\in\mathcal A,\quad
			\mathbf{x}^{\min}\le\mathbf{x}\le\mathbf{x}^{\max}\}.
		\end{aligned}
		\label{eq:X_v3}
	\end{equation}
	
\noindent	The PV operating point $(V_{\mathrm{pv}},I_{\mathrm{pv}})$ is obtained from
	the calibrated two-diode model under MPPT decoupling. The principal DC-side
	relations are
	\begin{subequations}\label{eq:pv_energy_chain}
		\begin{align}
			L_{\mathrm{dc}}
			&=I_{\mathrm{pv}}^2
			\frac{\rho_{\mathrm{dc}}\ell_{\mathrm{dc}}}{A_{\mathrm{dc}}},
			&
			V_{\mathrm{dc,inv}}
			&=V_{\mathrm{pv}}-I_{\mathrm{pv}}R_{\mathrm{dc}},\\
			\kappa_c
			&=\min\!\left(1,\frac{I_c}{I_{\mathrm{pv}}}\right),
			&
			P_{\mathrm{dc,in}}
			&=\kappa_c P_{\mathrm{dc}},\\
			E_{\mathrm{curt}}
			&=(1-\kappa_c)P_{\mathrm{dc}},
			&
			P_{\mathrm{PV}}^{\mathrm{AC}}
			&=\eta_{\mathrm{inv}}P_{\mathrm{dc,in}} .
		\end{align}
	\end{subequations}
	
\noindent	The feeder is represented by the full nonlinear balanced AC power-flow
	equations,
	\begin{subequations}
		\begin{align}
			P_i^{\mathrm{inj}}
			&=
			|V_i|\sum_j |V_j|
			\left[
			G_{ij}\cos(\delta_i-\delta_j)
			+B_{ij}\sin(\delta_i-\delta_j)
			\right],\\
			Q_i^{\mathrm{inj}}
			&=
			|V_i|\sum_j |V_j|
			\left[
			G_{ij}\sin(\delta_i-\delta_j)
			-B_{ij}\cos(\delta_i-\delta_j)
			\right].
		\end{align}
		\label{eq:acpf_v3}
	\end{subequations}
	
\noindent	Power delivered at the PCC is
	\begin{equation}
		P_{\mathrm{PCC}}(\mathbf{x},\xi)
		=
		P_{\mathrm{PV}}^{\mathrm{AC}}(\mathbf{x},\xi)
		-
		T_{\mathrm{ac,cable}}(\mathbf{x},\xi),
		\label{eq:pcc_power}
	\end{equation}
	\noindent and is injected into the selected bus before the backward--forward-sweep
	solution. Each scenario must satisfy the physical operating constraints
	\begin{equation}
		\begin{aligned}
			&V_i^{\min}\le |V_i|\le V_i^{\max},\qquad
			V_{\mathrm{dc}}^{\min}\le V_{\mathrm{dc,inv}}\le V_{\mathrm{dc}}^{\max},\\
			&S_{\mathrm{PV}}^{\mathrm{AC}}\le S_c,\qquad
			T_{\mathrm{pv}}\le T_{\mathrm{pv}}^{\max},\qquad
			|I_{ij}|\le I_{ij}^{\max}.
		\end{aligned}
		\label{eq:feasibility_v3}
	\end{equation}
	The chance-constrained safety event uses power-flow validity, bus-voltage
	limits, the PV-current cap, and inverter-power compliance; the remaining
	limits serve as scenario-level physical-validity diagnostics.
	
	Total operational loss is
	\begin{equation}
		\begin{aligned}
			T_L={}&T_{\mathrm{dc,cable}}
			+T_{\mathrm{dc,cond}}
			+T_{\mathrm{curt}}^{I}
			+T_{\mathrm{inv}}
			+T_{\mathrm{ac,cond}}\\
			&+T_{\mathrm{curt}}^{P}
			+T_{\mathrm{ac,cable}}
			+T_{\mathrm{network}},
		\end{aligned}
		\label{eq:total_loss_v3}
	\end{equation}
	where curtailment denotes foregone rather than dissipated power. The
	dissipative subset is\vspace{-7pt}
	\begin{equation}
		T_{\mathrm{diss}}
		=
		T_{\mathrm{dc,cable}}
		+T_{\mathrm{dc,cond}}
		+T_{\mathrm{inv}}
		+T_{\mathrm{ac,cond}}
		+T_{\mathrm{ac,cable}}
		+T_{\mathrm{network}} .
		\label{eq:dissipative_subset}
	\end{equation}
	
\noindent	PV penetration is evaluated scenario-wise as \vspace{-7pt}
	\begin{equation}
		\Pi(\mathbf{x},\xi)
		=
		100\frac{P_{\mathrm{PCC}}(\mathbf{x},\xi)}{P_D(\xi)},
		\label{eq:instant_penetration}
	\end{equation}
\noindent	with planning objective
	\begin{equation}
		\Pi(\mathbf{x})
		=
		\mathbb E_{\xi}
		\!\left[
		100\frac{P_{\mathrm{PCC}}(\mathbf{x},\xi)}{P_D(\xi)}
		\right].
		\label{eq:penetration_v3}
	\end{equation}
\noindent	Thus, penetration is normalized by scenario-dependent feeder demand rather
	than a fixed historical peak or installed-capacity denominator.

	\subsection{Stochastic Formulation}
	\label{sec:smop_v2}
	
\noindent	With $\mathbf{F}(\mathbf{x}) = \bigl[\mathbb{E}_\xi(T_L(\mathbf{x}, \xi)),\; -\Pi(\mathbf{x})\bigr]$ and the joint operational-safety event
	\begin{equation}
		\resizebox{0.97\linewidth}{!}{$
			\begin{aligned}
				\mathcal{S}(\mathbf{x},\xi)
				={}&
				\{\text{HF valid}\}
				\cap
				\{\text{PF converged}\}\cap
				\left\{
				V_i^{\min}
				\le V_i(\mathbf{x},\xi)
				\le V_i^{\max},
				\ \forall i
				\right\} \\[-1mm]
				&\cap
				\left\{
				I_{\mathrm{pv}}(\mathbf{x},\xi)
				\le r_{\max} I_c
				\right\}
				\cap
				\left\{
				S_{\mathrm{PV}}^{\mathrm{AC}}(\mathbf{x},\xi)
				\le S_c
				\right\},
			\end{aligned}
			$}
		\label{eq:joint_safety_event}
	\end{equation}
\noindent	the design problem is \vspace{-5pt}
	\begin{equation}
		\min_{\mathbf{x} \in \mathcal{X}} \quad \mathbf{F}(\mathbf{x}) \qquad \text{s.t.} \qquad \Pr_\xi\bigl(\mathcal{S}(\mathbf{x},\xi)\bigr) \ge 1-\alpha_{\mathrm{safe}},
		\label{eq:chance_constrained_v3}
	\end{equation}
	with $\Pr_\xi(\cdot)$ the true probability under the scenario distribution and $\alpha_{\mathrm{safe}}=0.10$ (a required reliability of 90\%); the estimator used to certify this against finite Monte Carlo evidence, $\underline{\Pr}_N(\cdot)$ (Eq.~\eqref{eq:confidence_bound_v3}), is constructed below and kept notationally distinct from the true probability it estimates. Every scenario contributes exactly one Bernoulli outcome --- completely safe, or not --- rather than one per constraint type: for any decomposition $\mathcal{S}=\bigcap_k\mathcal{S}_k$, $\Pr(\mathcal{S})\le\min_k\Pr(\mathcal{S}_k)$, so $\Pr(\mathcal{S})\ge0.90$ already forces every marginal event above 90\%, while the converse does not hold.
	
	The Pareto set
	\begin{equation}
		\resizebox{0.96\linewidth}{!}{$\mathcal{P}^{\star} = \Bigl\{\mathbf{x}\in\mathcal{X}_{\mathrm{feas}} : \nexists\,\mathbf{x}'\in\mathcal{X}_{\mathrm{feas}}\ \text{s.t.}\ \mathbf{F}(\mathbf{x}')\preceq\mathbf{F}(\mathbf{x}),\ \mathbf{F}(\mathbf{x}')\neq\mathbf{F}(\mathbf{x})\Bigr\}$}
		\label{eq:pareto_set_v3}
	\end{equation}
	gives the primary output $\Pi^{\star} = \{\Pi(\mathbf{x}) : \mathbf{x}\in\mathcal{P}^{\star}\}$, evaluated only on the independently HF-certified candidate set: the executed chronology is search $\to$ periodic enrichment $\to$ independent HF certification $\to$ HF-certified $\mathcal{P}^{\star}$ $\to$ TOPSIS selection (Algorithm~\ref{alg:enrichment}), applying TOPSIS (Section~\ref{sec:topsis}) to this certified set only.
	
\noindent	Four evaluation budgets are kept notationally and operationally distinct: $N_{\mathrm{ref}}=24$ (a reference/SAA convergence diagnostic, Section~\ref{sec:results_saa}); $N_{\mathrm{gen}}=60$ (the per-generation search budget, resampled every 20 generations); $N_{\mathrm{scr}}=100$ (the common screening batch, Section~\ref{sec:results_operational_selection}); and $N_{\mathrm{cert}}=500$ (independent certification, Section~\ref{sec:oos_verification}). The surrogate's training-set construction uses a separate budget, $N_{\mathrm{train}}$ (Section~\ref{sec:pins_v2}). Expectations use the standard SAA estimator\vspace{-10pt}
	\begin{equation}
		\widehat{\mathbb{E}}_N\bigl[F(\mathbf{x}, \xi)\bigr] = \frac{1}{N} \sum_{\nu=1}^{N} F(\mathbf{x}, \xi_\nu).
		\label{eq:saa_expect_v3}
	\end{equation}
	Two evaluations of \eqref{eq:chance_constrained_v3} are used at two stages and are deliberately not the same quantity. During search and screening, the raw empirical safe-outcome fraction is checked directly, with no confidence-interval discount,\vspace{-10pt}
	\begin{equation}
		\widehat{p}_N(\mathbf{x}) = \frac{1}{N}\sum_{\nu=1}^{N}\mathbb{1}\bigl[\mathcal{S}(\mathbf{x},\xi_\nu)\bigr], \qquad \widehat{p}_N(\mathbf{x}) \ge 1-\alpha_{\mathrm{safe}},
		\label{eq:search_fraction_v3}
	\end{equation}
	kept cheap precisely because nothing certified is claimed from it. Only at the independent $N_{\mathrm{cert}}=500$ stage is a one-sided lower confidence bound applied: the exact Clopper--Pearson bound $p^{L}_{\mathrm{CP},0.95}(\mathbf{x})$, at $1-\gamma=0.95$, used directly since it is exact for any sample size. The normal approximation,\vspace{-10pt}
	\begin{equation}
		\underline{\Pr}_N\bigl(\mathcal{S}(\mathbf{x},\xi)\bigr) = \widehat{p}_N(\mathbf{x}) - z_\gamma \sqrt{\frac{\widehat{p}_N(\mathbf{x})\bigl(1-\widehat{p}_N(\mathbf{x})\bigr)}{N}},
		\label{eq:confidence_bound_v3}
	\end{equation}
	is included for reference only and is not used for certification. A result reported against $N_{\mathrm{cert}}$ alone is described as certified; a result against $N_{\mathrm{scr}}$ is search-stage screened. Substituting \eqref{eq:search_fraction_v3} into \eqref{eq:chance_constrained_v3} gives the problem actually solved during the reported search,\vspace{-10pt}
	\begin{equation}
		\resizebox{0.95\linewidth}{!}{$\displaystyle\min_{\mathbf{x} \in \mathcal{X}} \left[ \frac{1}{N_{\mathrm{gen}}} \sum_{\nu=1}^{N_{\mathrm{gen}}} T_L(\mathbf{x}, \xi_\nu), \; -\,\Pi(\mathbf{x}) \right] \ \ \text{s.t.} \ \ \widehat{p}_{N_{\mathrm{gen}}}(\mathbf{x}) \ge 1-\alpha_{\mathrm{safe}},$}
		\label{eq:saa_moop_v3}
	\end{equation}
	resampled every 20 generations. $N_{\mathrm{gen}}=60$ is not arbitrary: the $N_{\mathrm{ref}}$ diagnostic first shows $\widehat{\mathbb{E}}_{N}[T_L]$ and $\Pi(\mathbf{x})$ stabilise within tolerance by $N_{\mathrm{ref}}=24$ (Section~\ref{sec:results_saa}); the executed search then evaluates at the larger, independently fixed $N_{\mathrm{gen}}=60$. Scenarios are built by empirical resampling of complete observed weather states --- irradiance, solar geometry, albedo, and ambient temperature drawn together, preserving within-weather dependence as recorded --- combined with an independently sampled demand multiplier, $m_L\sim\mathrm{Lognormal}(\mu_L,\sigma_L^2)$ truncated to $[0.70,1.30]$, $\sigma_L=0.10$, $\mu_L=-\tfrac12\sigma_L^2$ so $\mathbb{E}[m_L]\approx1$; the scenario-independence argument underlying the confidence bound above is given in Appendix~\ref{appscenario}.
	
	\subsection{Neural Surrogate for Stochastic Evaluation}
	\label{sec:pins_v2}
	\label{sec:baselines}
	
\begin{figure}
		\centering
		\includegraphics[width=0.99\columnwidth, height=5.9cm, keepaspectratio]{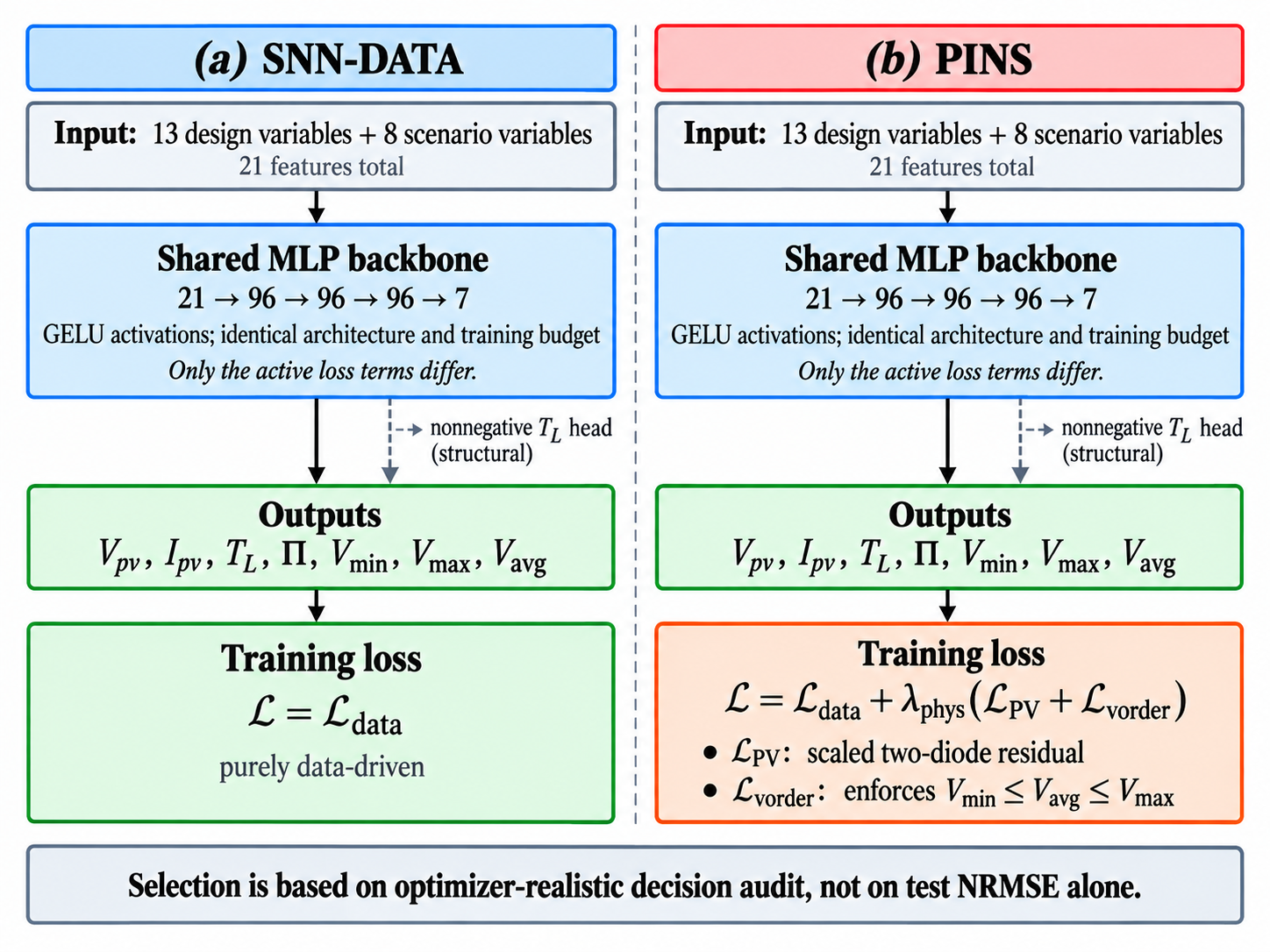}
		\caption{Matched SNN-DATA and PINS architectures; PINS adds physics-informed loss terms, while model selection is based on operational high-fidelity auditing.}
		\label{fig:current_architecture}
	\end{figure}

	The surrogate replaces a per-scenario two-diode-plus-power-flow solve with a single forward pass, predicting the decision-relevant physical state directly rather than the raw bus-level state vector, via a shared trunk with GELU activations at every hidden layer:
	{\small
		\begin{subequations}\label{eq:surrogate_block}
			\begin{align}
				\mathcal{F}&: (\mathbf{x},\xi)\mapsto\mathbf{s}=[V_{\mathrm{pv}},I_{\mathrm{pv}},T_L,\Pi,V_{\min},V_{\max},V_{\mathrm{avg}}]^{T}\in\mathbb{R}^{7}, \notag\\
				&\hspace{1.6em} \widehat{\mathbf{s}}=\mathcal{F}_\theta(\mathbf{z})\approx\mathcal{F}(\mathbf{x},\xi), \label{eq:surrogate_mapping}\\
				\widehat{\mathbf{s}} &= \mathbf{W}_4\,\sigma\!\bigl(\mathbf{W}_3\,\sigma(\mathbf{W}_2\,\sigma(\mathbf{W}_1\mathbf{z}+\mathbf{b}_1)+\mathbf{b}_2)+\mathbf{b}_3\bigr)+\mathbf{b}_4, \notag\\
				&\hspace{3.4em} 21\!\to\!96\!\to\!96\!\to\!96\!\to\!7, \label{eq:architecture_v3}\\
				\sigma(u) &= u\,\Phi(u), \qquad \Phi(u)=\tfrac12\bigl[1+\operatorname{erf}(u/\sqrt2)\bigr]. \label{eq:gelu_v3}
			\end{align}
		\end{subequations}
	}
\noindent	Here $\mathbf{z}=[\mathbf{x},\boldsymbol\xi]\in\mathbb{R}^{21}$ concatenates the thirteen design variables of \eqref{eq:decision_vector} with eight scenario variables (irradiance, ambient temperature, solar geometry, albedo, and the demand multiplier of Section~\ref{sec:smop_v2}); predicting the five decision-relevant outputs directly, rather than reconstructing them from raw bus voltages and angles, avoids separate loss/penetration heads that individually score well but silently disagree on the same operating point. Only $\widehat{T}_L$ passes through a softplus link, so it is structurally nonnegative regardless of $\theta$. Architecture, widths, and training budget are identical across both compared variants; only the active loss terms differ. The data-only variant (SNN-DATA) minimises $\mathcal{L}_{\mathrm{data}}$ alone; the physics-penalised variant (PINS) uses the identical architecture, data, and budget with two additional residual penalties evaluated on the surrogate's own predictions \vspace{-10pt}:
	\begin{subequations}\label{eq:loss_block}
		\begin{align}
			\mathcal{L}_{\mathrm{data}}
			&=
			\frac{1}{N_d}\sum_{k=1}^{N_d}
			\left\|
			\widehat{\mathbf{s}}^{(k)}-\mathbf{s}^{(k)}
			\right\|_2^2,
			\label{eq:data_loss_v3}
			\\
			\mathcal{L}_{\mathrm{PINS}}(\theta)
			&=
			\mathcal{L}_{\mathrm{data}}
			+\lambda_{\mathrm{phys}}
			\left(
			\mathcal{L}_{\mathrm{PV}}
			+\mathcal{L}_{\mathrm{vorder}}
			\right),
			\label{eq:complete_loss_v3}
			\\
			\mathcal{L}_{\mathrm{PV}}
			&=
			\frac{1}{N_d}\sum_{k=1}^{N_d}
			\left[
			\operatorname{asinh}
			\left(
			\frac{\mathcal{R}_I^{(k)}}{s_I+\epsilon}
			\right)
			\right]^2,
			\label{eq:pv_phys_loss}
			\\
			r_{\mathrm{vorder}}
			&=
			\max\!\left(0,\widehat V_{\min}-\widehat V_{\mathrm{avg}}\right)
			\nonumber\\
			&\quad+
			\max\!\left(0,\widehat V_{\mathrm{avg}}-\widehat V_{\max}\right).
			\label{eq:voltage_coupling_v3}
		\end{align}
	\end{subequations}
	$\mathcal{L}_{\mathrm{PV}}$ evaluates the two-diode current residual $\mathcal{R}_I$ (Appendix~\ref{apptd}) on the surrogate's own predicted $\widehat V_{\mathrm{pv}},\widehat I_{\mathrm{pv}}$, scaled by a fixed training-set-level scale $s_I$; $\mathcal{L}_{\mathrm{vorder}}$ is a physically necessary ordering-consistency penalty, zero exactly when $\widehat V_{\min}\le\widehat V_{\mathrm{avg}}\le\widehat V_{\max}$ and linear beyond it. Because the two-diode residual is exponential in $V_{\mathrm{pv}}, I_{\mathrm{pv}}$, an unbounded squared penalty can destabilise training even at small $\lambda_{\mathrm{phys}}$; $\mathcal{L}_{\mathrm{PV}}$ instead squares an $\operatorname{asinh}$-compressed residual, combined with gradient-norm clipping and $\lambda_{\mathrm{phys}}$ chosen by grid search against $\mathcal{D}_{\mathrm{val}}$ RMSE. SNN-DATA is exactly this architecture and protocol with $\lambda_{\mathrm{phys}}=0$, making the comparison below a controlled ablation. Fig.~\ref{fig:current_architecture} contrasts the two variants.
	
\noindent	
	Training, validation, and test sets are separated by design, with normalization fitted only on the training set: $\mathcal{D}_{\mathrm{train}}=\{(\mathbf{x}^{(k)},\xi^{(k)},\mathbf{s}^{(k)})\}_{k=1}^{N_d}$ (Latin Hypercube sampling with the $N_{\mathrm{train}}$-scenario procedure), $\mathcal{D}_{\mathrm{val}}$ (selection and physics-weight search), $\mathcal{D}_{\mathrm{test}}$ (held out before any active learning, used once), and $\mathcal{D}_{\mathrm{enrich}}$ (grown by periodic enrichment, rejoins $\mathcal{D}_{\mathrm{train}}$ only)-- identically for both variants. Both surrogates undergo a common independent HF operational audit on the $N_{\mathrm{scr}}=100$ screening batch, and the model satisfying the pre-specified decision-validity gate --- not necessarily the lower-error one --- is retained to drive the reported stochastic optimization (Section~\ref{sec:results_operational_selection}); the rejected variant generates none of the reported Pareto fronts. Neither variant is load-bearing to the paper's central claim: the enrichment loop of Section~\ref{sec:enrichment_v2} audits either variant's errors identically. Two independently trained instances of the deployed variant, $\theta_1^{(0)}, \theta_2^{(0)}$, serve as the disagreement heuristic below, updated to $\theta_1^{(r)},\theta_2^{(r)}$ at each accepted enrichment round.
	
	\subsection{Periodic Decision-Gated High-Fidelity Enrichment and Rollback}
	\label{sec:adaptive_v2}
	\label{sec:enrichment_v2}
	
\noindent	This is the paper's primary methodological contribution. As optimization progresses, the queried design distribution concentrates near Pareto and active-constraint regions, where small surrogate errors can alter feasibility or dominance. At scheduled enrichment generations $g\in\mathcal{G}_{\mathrm{enrich}}$ (round $r$), three candidate regions are drawn from the current population and re-evaluated against $\mathcal{F}$, with a fixed per-round quota (30 Pareto + 45 boundary + 45 disagreement candidates, deduplicated): \textbf{Pareto} $\mathcal{D}_{\mathrm{Pareto}}^{(r)}$, the current front, expected to carry the highest discrepancy since the search exploits it most aggressively; \textbf{boundary} $\mathcal{D}_{\mathrm{boundary}}^{(r)}$, proximity to a binding limit; and \textbf{disagreement} $\mathcal{D}_{\mathrm{disagreement}}^{(r)}$, divergence between the two independently trained surrogates:\vspace{-10pt}
	\begin{equation}
		\mathcal{D}_{\mathrm{new}}^{(r)} = \mathcal{D}_{\mathrm{Pareto}}^{(r)} \cup \mathcal{D}_{\mathrm{boundary}}^{(r)} \cup \mathcal{D}_{\mathrm{disagreement}}^{(r)}.
		\label{eq:periodic_acquisition}
	\end{equation}
	Using only the quantities the surrogate predicts (Eq.~\eqref{eq:surrogate_mapping}), boundary proximity and disagreement are
	{\small
		\begin{subequations}\label{eq:acquisition_block}
			\begin{align}
				d^{(r)}(\mathbf{x},\xi_\nu) &= \min\bigl\{\widehat V_{\min}-V^{\min},\; V^{\max}-\widehat V_{\max},\; r_{\max}I_c-\widehat I_{\mathrm{pv}}\bigr\}, \notag\\
				d^{(r)}(\mathbf{x}) &= \mathrm{quantile}_\alpha\bigl(\{d^{(r)}(\mathbf{x},\xi_\nu)\}_\nu\bigr), \label{eq:margin_v3}\\
				u^{(r)}(\mathbf{x},\xi) &= \frac{\|\widehat{\mathbf{s}}_{\theta_1^{(r)}}(\mathbf{x},\xi) - \widehat{\mathbf{s}}_{\theta_2^{(r)}}(\mathbf{x},\xi)\|_2}{s_{\mathbf{s}}+\epsilon}. \label{eq:disagreement_v3}
			\end{align}
		\end{subequations}
	}
	Small $d^{(r)}$ identifies constraint-critical designs; large $u^{(r)}$ identifies poorly resolved regions, as an acquisition heuristic only, with no probabilistic coverage guarantee claimed.
	
	Acquired candidates are evaluated with $\mathcal{F}$ and accumulated across rounds,
	\begin{equation}
		\mathcal{D}_{\mathrm{enrich}}^{(r)}
		=
		\mathcal{D}_{\mathrm{enrich}}^{(r-1)}
		\cup
		\mathcal{D}_{\mathrm{new}}^{(r)}.
		\label{eq:periodic_enrichment_update}
	\end{equation}
	The resulting HF labels give per-round relative discrepancies in loss and penetration, and a false-feasibility rate capturing the decision-critical failure mode in which a surrogate-feasible design is HF-infeasible:
	\begin{subequations}\label{eq:discrepancy_block}
		\begin{align}
			e_T^{(r)}(\mathbf{x})
			&=
			\frac{\left|T_L^{\mathcal F}(\mathbf{x})-\widehat T_L(\mathbf{x})\right|}
			{\max\!\left(\left|T_L^{\mathcal F}(\mathbf{x})\right|,\epsilon_0\right)},
			\label{eq:discrepancy_T_v3}
			\\
			e_{\Pi}^{(r)}(\mathbf{x})
			&=
			\frac{\left|\Pi^{\mathcal F}(\mathbf{x})-\widehat{\Pi}(\mathbf{x})\right|}
			{\max\!\left(\left|\Pi^{\mathcal F}(\mathbf{x})\right|,\epsilon_0\right)},
			\label{eq:discrepancy_v3}
			\\
			F_{\mathrm{false}}^{(r)}
			&=
			\frac{\left|\left\{\mathbf{x}\in\mathcal D_{\mathrm{new}}^{(r)}:\begin{array}{l}\mathbf{x}\ \text{surrogate-feasible},\\[-1mm]\mathbf{x}\ \text{HF-infeasible}\end{array}\right\}\right|}
			{\left|\left\{\mathbf{x}\in\mathcal D_{\mathrm{new}}^{(r)}:\mathbf{x}\ \text{surrogate-feasible}\right\}\right|}.
			\label{eq:ffalse_v3}
		\end{align}
	\end{subequations}
	A candidate update $\theta_{\mathrm{cand}}^{(r)}$ is retrained on $\mathcal{D}_{\mathrm{train}}\cup\mathcal{D}_{\mathrm{enrich}}^{(r)}$ at each round (grid search over correction strength) and accepted only if the prediction-error, false-feasibility, and decision-agreement gates all pass:
	\begin{equation}
		\theta^{(r)} =
		\begin{cases}
			\theta_{\mathrm{cand}}^{(r)}, & \mathcal{G}_{\mathrm{err}}^{(r)}\,\mathcal{G}_{\mathrm{FF}}^{(r)}\,\mathcal{G}_{\mathrm{DA}}^{(r)}=1,\\
			\theta^{(r-1)}, & \text{otherwise}.
		\end{cases}
		\label{eq:decision_gate_update}
	\end{equation}
	Enrichment is thus accepted for improved engineering decisions, not merely lower regression error; a failed update rolls back to $\theta^{(r-1)}$ while the search continues. Section~\ref{sec:results_enrichment} reports this targeted acquisition against a random-enrichment control and single-source variants at a matched per-round budget, isolating acquisition location from acquisition quantity.
	
	\begin{algorithm}
		\caption{Decision-gated surrogate-assisted optimization with periodic high-fidelity enrichment and rollback}
		\label{alg:enrichment}
		\begin{algorithmic}[1]
			\State \textbf{Select:} train $\mathcal{F}_{\theta_1^{(0)}},\mathcal{F}_{\theta_2^{(0)}}$ on $\mathcal{D}_{\mathrm{train}}$; fix $\mathcal{D}_{\mathrm{val}},\mathcal{D}_{\mathrm{test}}$; choose the operational surrogate by independent HF re-evaluation on the screening batch (Section~\ref{sec:results_operational_selection}).
			\State \textbf{Initialize:} mixed-integer MOEA population over $\mathcal{X}$; $g\gets0$; $r\gets0$; $\mathcal{D}_{\mathrm{enrich}}^{(0)}\gets\emptyset$.
			\While{not converged}
			\State evaluate the population via the current surrogate $\mathcal{F}_{\theta^{(r)}}$ (Eq.~\ref{eq:saa_moop_v3}, $N_{\mathrm{gen}}\!=\!60$, resampled every 20 generations); $g\gets g+1$.
			\If{$g\in\mathcal{G}_{\mathrm{enrich}}$}
			\State $r\gets r+1$.
			\State \textbf{Acquire:} form $\mathcal{D}_{\mathrm{new}}^{(r)}$ (Eq.~\ref{eq:periodic_acquisition}, Eqs.~\ref{eq:margin_v3}--\ref{eq:disagreement_v3}) from the current population; evaluate with $\mathcal{F}$; update $\mathcal{D}_{\mathrm{enrich}}^{(r)}$ (Eq.~\ref{eq:periodic_enrichment_update}); compute $e_T^{(r)},e_\Pi^{(r)}$ (Eq.~\ref{eq:discrepancy_v3}), $F_{\mathrm{false}}^{(r)}$ (Eq.~\ref{eq:ffalse_v3}).
			\State retrain candidate $\theta_{\mathrm{cand}}^{(r)}$ on $\mathcal{D}_{\mathrm{train}}\!\cup\!\mathcal{D}_{\mathrm{enrich}}^{(r)}$; audit decision agreement against $\mathcal{F}$ on held-out paired designs.
			\State apply the decision gate (Eq.~\ref{eq:decision_gate_update}): \textbf{accept} $\theta^{(r)}\gets\theta_{\mathrm{cand}}^{(r)}$ if $\mathcal{G}_{\mathrm{err}}^{(r)}\mathcal{G}_{\mathrm{FF}}^{(r)}\mathcal{G}_{\mathrm{DA}}^{(r)}=1$; \textbf{else reject and roll back}: $\theta^{(r)}\gets\theta^{(r-1)}$.
			\EndIf
			\EndWhile
			\State \textbf{Certify:} re-verify the retained surrogate's $\widehat{\mathcal{P}}^{\star}$ against $\mathcal{F}$ on an independent $N_{\mathrm{cert}}$-scenario set (Section~\ref{sec:oos_verification}) $\to\mathcal{P}^{\star},\Pi^{\star}$.
			\State \textbf{Decide:} select one design via TOPSIS (Section~\ref{sec:topsis}) from the HF-certified $\mathcal{P}^{\star}$, where required.
		\end{algorithmic}
	\end{algorithm}
	
	\begin{figure*}
		\centering
		\includegraphics[width=0.7\textwidth]{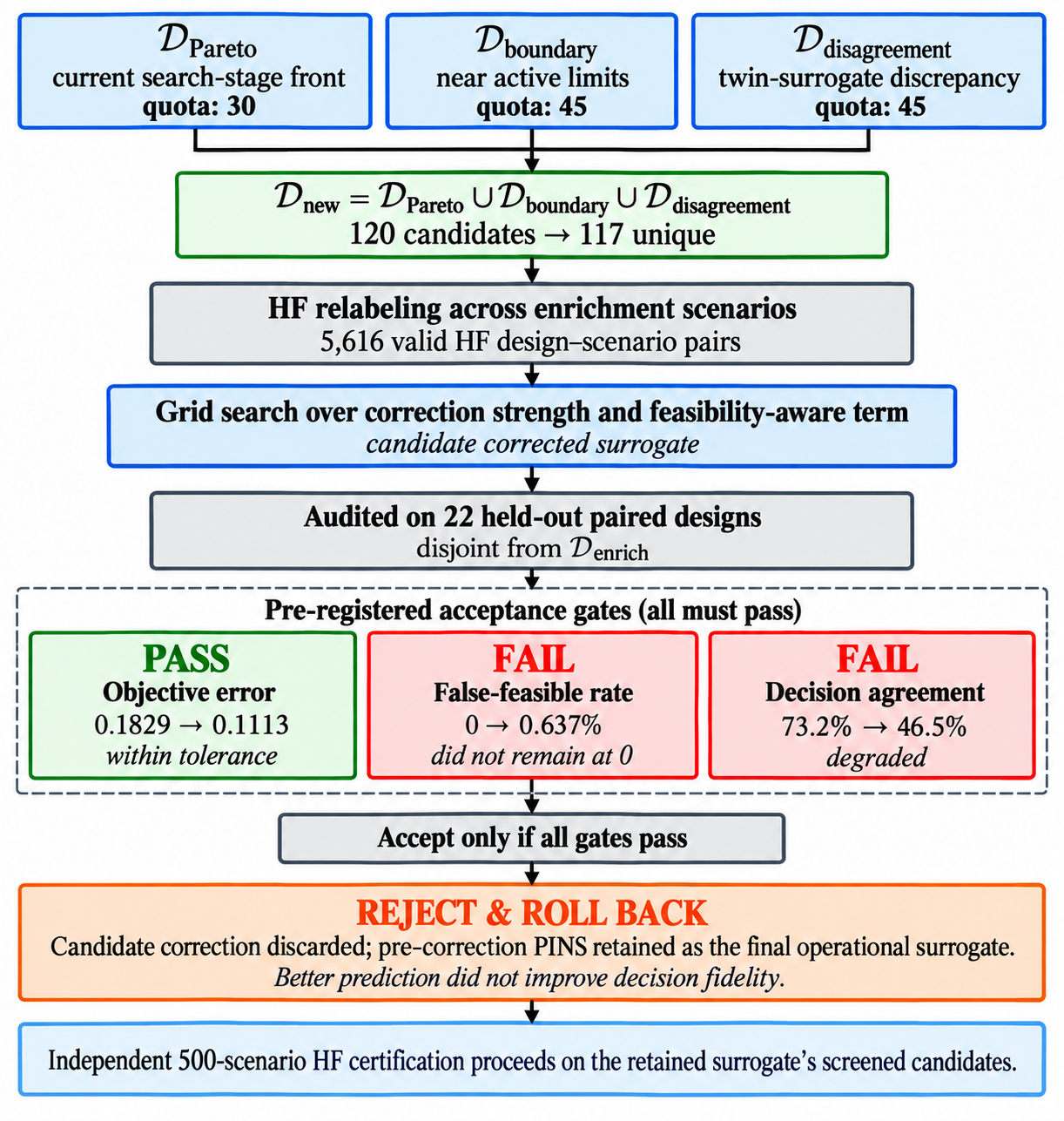}
		\caption{Executed periodic enrichment pipeline (Algorithm~\ref{alg:enrichment}): the audited round's acquisition and decision-gate outcome; full numerical results are reported in Section~\ref{sec:results_enrichment}.}
		\label{fig:current_correction_pipeline}
	\end{figure*}
	\subsection{Statistical Protocol and Validation}
	\label{sec:validation_protocol}
	
	\noindent
	Uncertainty is separated into sampling, surrogate, and search components:\vspace{-10pt}
	\begin{equation}
		\mathcal{E}
		=
		\left\{
		\mathcal{E}_{\mathrm{SAA}},
		\mathcal{E}_{\mathrm{sur}},
		\mathcal{E}_{\mathrm{search}}
		\right\},
		\qquad
		\mathcal{E}_{\mathrm{sur}}
		\equiv
		\{\Delta_f,\Delta_c,\mathrm{FF},\mathrm{FI}\},
		\label{eq:error_partition}
	\end{equation}
	where surrogate discrepancies are defined in
	Eqs.~\eqref{eq:discrepancy_v3}--\eqref{eq:ffalse_v3}, and search variability
	is estimated over $R=8$ independent seeds. The principal comparison is the
	recovery of HF-valid, independently certified Pareto designs relative to the
	HF computational cost,\vspace{-5pt}
	\begin{equation}
		\eta_{\mathrm{HF}}
		=
		\frac{N_{\mathrm{HF}}^{\mathrm{sur}}}
		{N_{\mathrm{HF}}^{\mathrm{direct}}},
		\qquad
		\eta_{\mathrm{HF}}<1.
		\label{eq:hf_efficiency}
	\end{equation}
	All physical solvers are validated before stochastic optimization, and all
	parameters defining $\mathcal X$ and $\mathcal F$ are fixed in the
	reproducibility record.
	
	\subsection{Independent High-Fidelity Verification}
	\label{sec:oos_verification}
	
	\noindent
	Final candidates are evaluated on an independent certification set\vspace{-10pt}
	\begin{equation}
		\mathcal D_{\mathrm{cert}}
		\cap
		\left(
		\mathcal D_{\mathrm{train}}
		\cup
		\mathcal D_{\mathrm{opt}}
		\cup
		\mathcal D_{\mathrm{enrich}}
		\right)
		=
		\varnothing,
		\qquad
		|\mathcal D_{\mathrm{cert}}|=N_{\mathrm{cert}} .
		\label{eq:cert_independence}
	\end{equation}
	
	For candidate $\mathbf{x}$, let
	\begin{equation}
		K(\mathbf{x})
		=
		\sum_{\nu=1}^{N_{\mathrm{cert}}}
		\mathbf{1}
		\!\left[
		\mathcal S(\mathbf{x},\xi_\nu)
		\right],
		\qquad
		\widehat p_{\mathrm{safe}}
		=
		\frac{K(\mathbf{x})}{N_{\mathrm{cert}}}.
		\label{eq:cert_successes}
	\end{equation}
	The exact one-sided Clopper--Pearson lower bound is
	\begin{equation}
		p_{\mathrm{CP},0.95}^{L}(\mathbf{x})
		=
		F_{\mathrm{Beta}}^{-1}
		\!\left(
		0.05;\,
		K,\,
		N_{\mathrm{cert}}-K+1
		\right),
		\label{eq:cp_lower}
	\end{equation}
	and certification requires
	\begin{equation}
		p_{\mathrm{CP},0.95}^{L}(\mathbf{x})
		\ge
		1-\alpha_{\mathrm{safe}} .
		\label{eq:cert_rule}
	\end{equation}
	
	All reported final quantities are recomputed by the HF model:
	\begin{equation}
		\left(
		\widehat E[T_L],\,
		\widehat E[\Pi],\,
		\widehat p_{\mathrm{safe}}
		\right)_{\mathrm{final}}
		=
		\left(
		\widehat E[T_L],\,
		\widehat E[\Pi],\,
		\widehat p_{\mathrm{safe}}
		\right)_{\mathrm{HF}} .
		\label{eq:hf_final_authority}
	\end{equation}
	Solver failure or non-finite output implies
	$\mathbf{1}[\mathcal S]=0$; repeated designs are cached and invalid training
	samples are excluded. Seeds, software versions, dataset sizes, checksums, and
	diagnostics are retained in the reproducibility manifest.

	%=====================================================================
	\section{Simulations, Analyses and Discussions of Results}
	\label{sec:sim_results_current}
	%=====================================================================
	
\noindent	This section follows the computational evidence chain established in the
	methodology: simulation-data and network audit, HF physical validation,
	stochastic adequacy, matched-surrogate assessment, operational HF selection,
	periodic HF enrichment, independent reliability certification, compromise
	selection, physical interpretation of the loss objective, and cross-feeder
	engineering transfer. This ordering deliberately separates three quantities
	that should not be conflated in surrogate-assisted planning:
	\emph{predictive accuracy}, \emph{decision reliability}, and
	\emph{independent engineering certification}.
	
	%---------------------------------------------------------------------
	\subsection{Simulation Configuration, Data Audit, and Reproducibility}
	\label{sec:results_data_audit}
	%---------------------------------------------------------------------
	
	\noindent The primary study uses a 34-bus radial distribution feeder on a 12.66-kV
	base with an admissible voltage range of 0.95--1.05~pu. A scenario is
	operationally safe only when the HF evaluation is valid, the power flow
	converges, bus voltages remain within limits, and the PV-current and
	inverter-power constraints are satisfied simultaneously. The planning loss
	cap is 10\% of the nominal 4{,}636.5~kW feeder demand, or 463.65~kW.
	The complete 13-dimensional design domain was audited before optimization,
	and the declared bounds retain the intended catalogue and continuous-variable\\
	ranges.
	
\begin{figure}
		\centering
		\includegraphics[width=\linewidth,height=4.0cm]{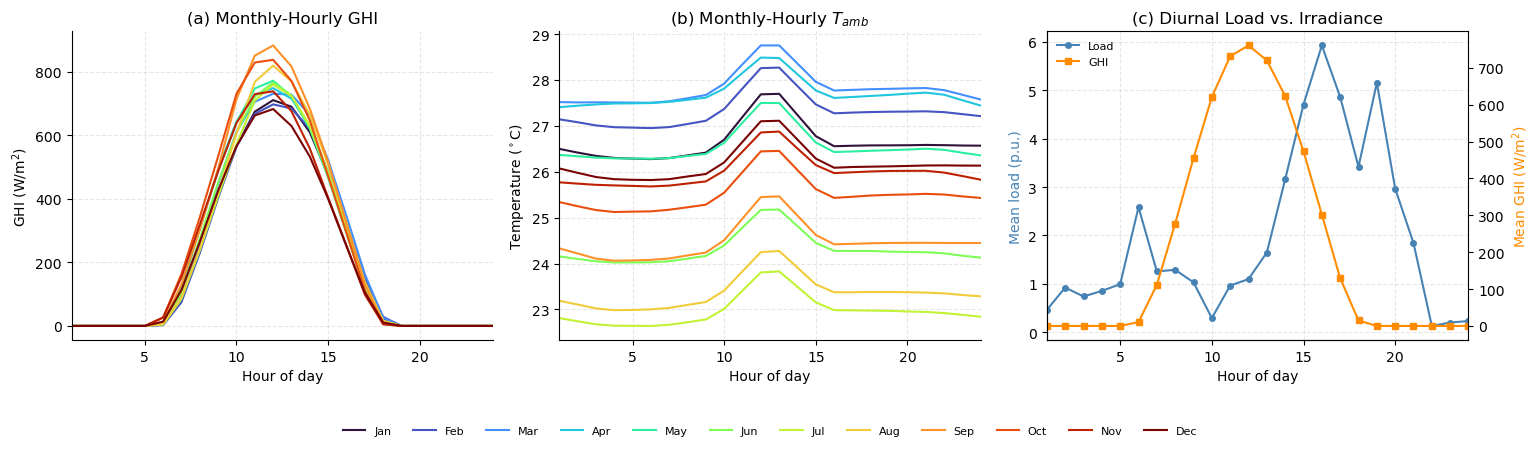}
		\caption{Historical source-data climatology: (a) GHI, (b) ambient
			temperature, and (c) mean diurnal demand and irradiance.}
		\label{fig:current_climatology}
	\end{figure}

\noindent	The historical archive contains irradiance, ambient temperature, solar
	geometry, albedo, and demand information. Data-quality screening removed
	invalid weather states and duplicate timestamps before scenario generation.	
	Figure~\ref{fig:current_climatology} summarizes the source climatology.
	Irradiance follows the expected daylight cycle, whereas demand remains
	substantial outside the strongest solar-production interval and exhibits a
	secondary evening peak. The resulting temporal mismatch motivates explicit
	demand uncertainty rather than deterministic coincidence between PV
	production and load.
	
\begin{figure}
		\centering
		\includegraphics[width=1.1\linewidth]{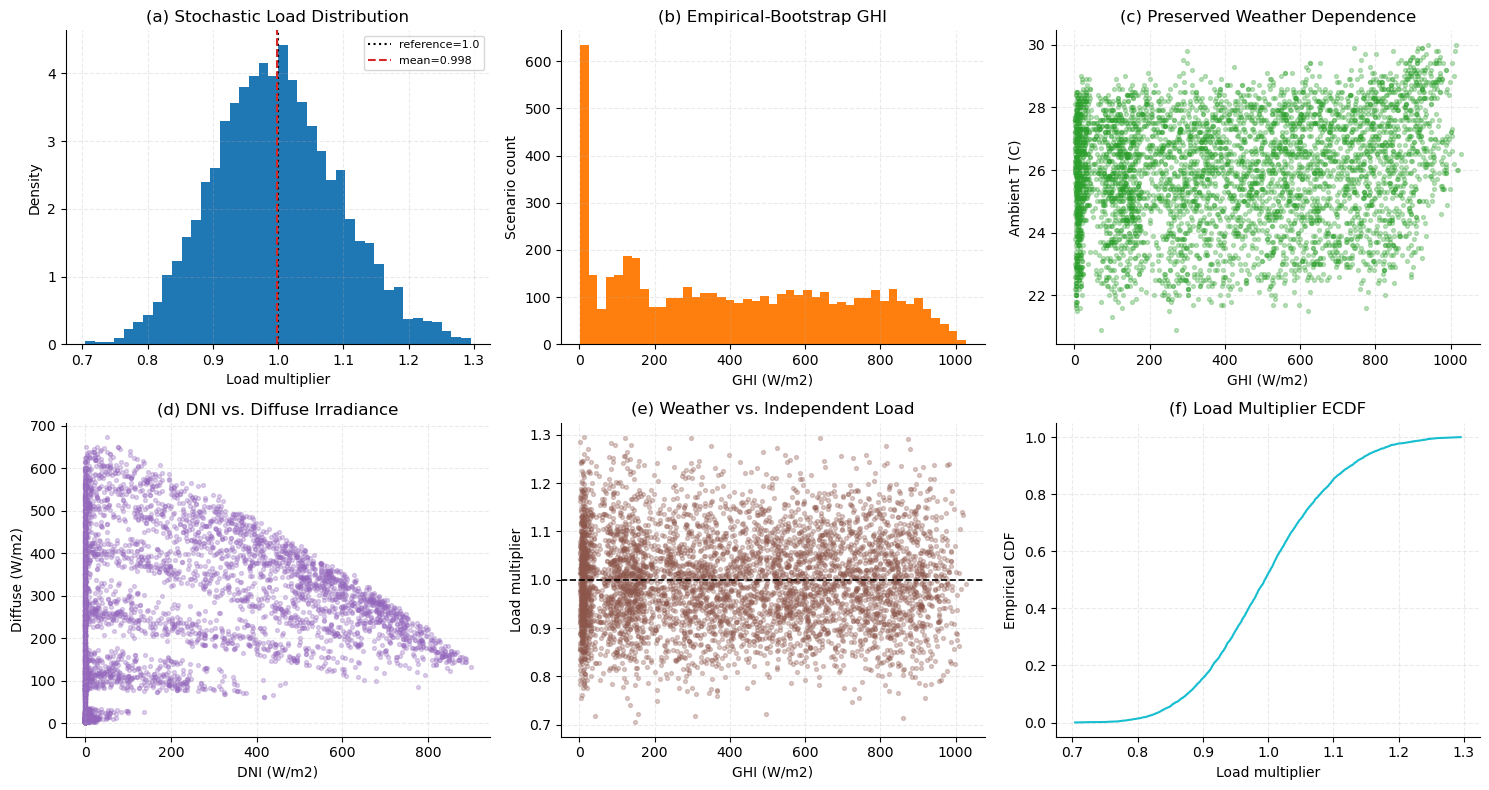}
		\caption{Audit of 250 weather--demand scenarios used to assess the
			SAA sampling distribution.}
		\label{fig:current_scenarios}
	\end{figure}

	\noindent Stochastic scenarios combine one complete empirical weather row with an
	independently sampled mean-one truncated-lognormal demand multiplier
	($\sigma_L=0.10$, support $[0.70,1.30]$). Complete-row resampling preserves
	the observed dependence among irradiance, temperature, solar geometry, and
	albedo without fitting a separate parametric weather distribution, whereas
	demand uncertainty is introduced independently. The historical load profile
	in Fig.~\ref{fig:current_climatology} is therefore descriptive context rather
	than the scenario-generation mechanism itself. The resulting stochastic space was audited using 250 weather--demand
	scenarios (Fig.~\ref{fig:current_scenarios}). All 250 produced valid HF
	evaluations for the audited design; 98.0\% satisfied the joint operational
	safety event, while 85.2\% simultaneously occupied the prescribed
	penetration and loss windows. Hence, the scenario distribution contains both
	nominal and stressed operating states rather than being concentrated around
	an artificially easy operating regime.
	
\begin{figure}
		\centering
		\includegraphics[width=1.0\linewidth,height=4.0cm]{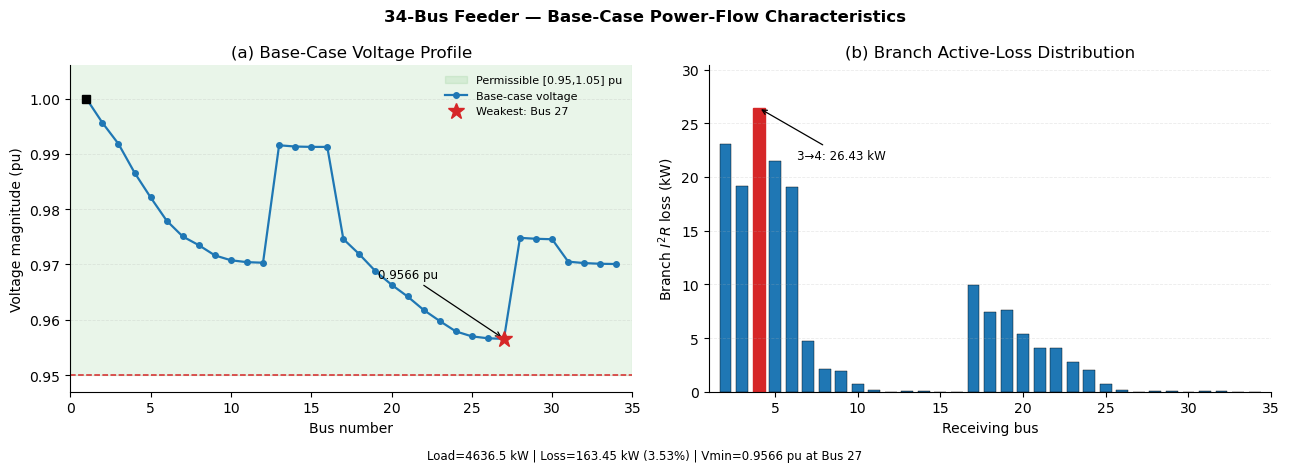}
		\caption{No-PV 34-bus baseline: (a) voltage profile and
			(b) branch-loss distribution.}
		\label{fig:current_voltage}
	\end{figure}

\noindent	The no-PV feeder was also independently verified before optimization.
	It carries 4{,}636.5~kW active and 2{,}873.5~kVAr reactive demand, with
	network $I^2R$ loss of 163.45~kW (3.53\%). The minimum voltage is
	0.9566~pu at bus 27, above the 0.95~pu lower limit
	(Fig.~\ref{fig:current_voltage}). Thus, the optimization begins from a
	converged and voltage-compliant reference network rather than from an already
	infeasible operating state.
	
\noindent	The largest baseline branch loss, 26.43~kW, occurs on section
	$3\!\to\!4$, reflecting the concentration of upstream feeder losses and
	providing an early indication that the electrical location of PV injection
	can materially affect network performance.

%---------------------------------------------------------------------
\subsection{High-Fidelity Physical-Model Validation}
\label{sec:results_physics}
\label{sec:phys_validation}
%---------------------------------------------------------------------

\begin{figure}
	\centering
	\includegraphics[width=\linewidth,height=4.0cm]{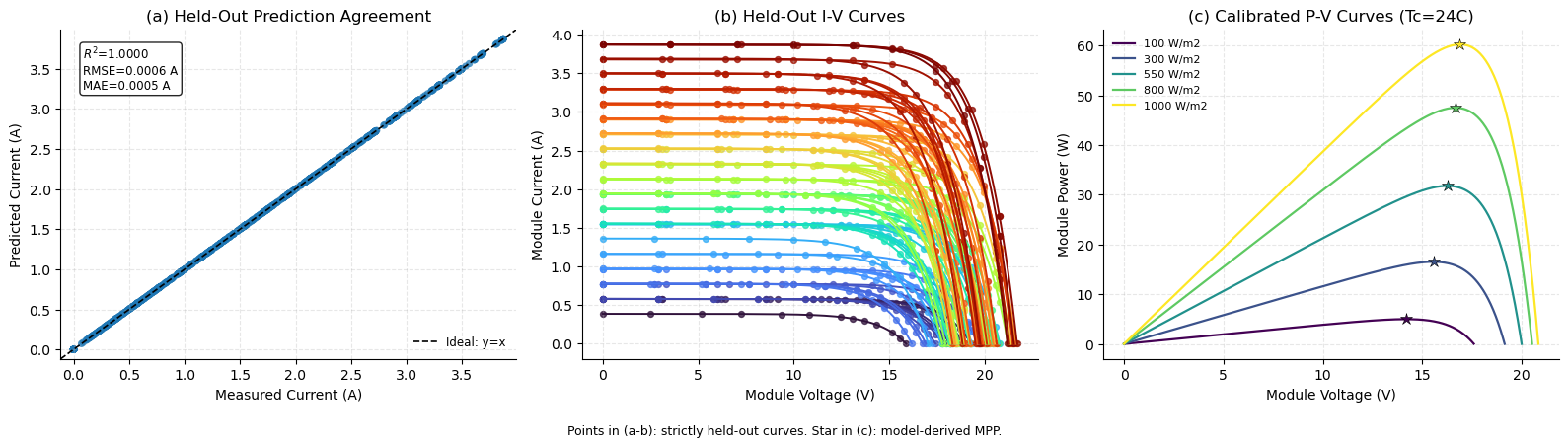}
	\caption{Curve-disjoint validation of the calibrated two-diode PV model:
		(a) held-out current agreement, (b) withheld I--V curves, and
		(c) P--V characteristics with model-derived maximum-power points.}
	\label{fig:current_pv_validation}
\end{figure}

\noindent
The PV reference model was calibrated using the publicly available MSX-60
characterization dataset of Harrison et. al
\cite{HarrisonAlombah2022PVData,HarrisonAlombah2023PVE}, comprising
399 irradiance--temperature conditions and 27 I--V/P--V variables. A
curve-disjoint split used 319 curves for calibration and 80 for independent
validation. The two-diode model was solved by damped Newton--Raphson with a
post-convergence residual check and achieved
$R^2=0.99999966$, RMSE $=0.615$~mA, and MAE $=0.504$~mA on the held-out
curves; the nearest STC maximum-power point differed by only $-0.013\%$
(Fig.~\ref{fig:current_pv_validation}). Two of the eight fitted parameters
reached their imposed bounds, indicating strong predictive agreement but
limited identifiability for those parameters.

\noindent
These results, together with the feeder validation, establish the HF simulator
as the reference for surrogate assessment, optimization decisions, and final
certification.
	%---------------------------------------------------------------------
	\subsection{Stochastic Adequacy and SAA Budget}
	\label{sec:results_saa}
	%---------------------------------------------------------------------
	
	The planning problem is a stochastic, chance-constrained multiobjective
	optimization that minimizes expected operational loss while maximizing
	expected PV penetration, subject to a joint operational-safety probability
	of at least 0.90. The demand multiplier scales the feeder's nominal active
	and reactive loads simultaneously, while weather variables within each
	scenario retain the dependence observed in the sampled historical row.
	Weather and demand are not jointly bootstrapped; consequently, the executed
	sampler preserves within-weather dependence but does not impose historical
	weather--demand temporal dependence scenario by scenario.
	A nested convergence study evaluated common scenario prefixes of
	$N=12,24,48,$ and $96$. Objective variability decreased systematically as
	$N$ increased, supporting $N_{\mathrm{ref}}=24$ as an evidence-checked
	diagnostic reference size. The actual optimization uses the larger
	$N_{\mathrm{gen}}=60$ scenario budget per generation, refreshed every
	20 generations, whereas the final reliability calculation uses an
	independent $N_{\mathrm{cert}}=500$ HF sample. These budgets therefore serve
	distinct purposes: convergence diagnosis, stochastic search, and statistical
	certification, respectively.
	Across the five audited designs, the standard-deviation decay was
	approximately $N^{-0.62}$ for $T_L$ and $N^{-0.64}$ for $\Pi$; at
	$N=24$, the panel-average relative standard errors were approximately
	2.17\% and 3.08\%, respectively.
	
	%---------------------------------------------------------------------
	\subsection{Matched Surrogates: Predictive Performance and Robustness}
	\label{sec:results_surrogates}
	\label{sec:results_ablation}
	%---------------------------------------------------------------------
	
\begin{figure}
		\centering
		\includegraphics[
		width=1.08\linewidth,
		height=4.2cm]{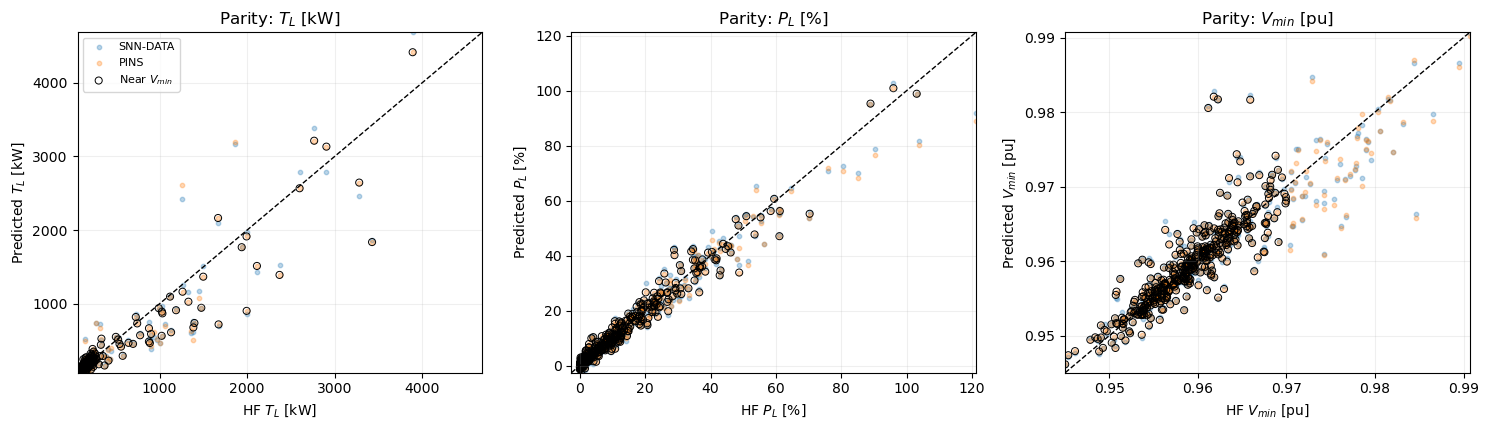}
\caption{Held-out parity for total loss, penetration, and minimum voltage on the 450-pair test set; circled points indicate samples near the $V_{\min}$ limit.}
		\label{fig:current_parity}
	\end{figure}

\noindent	SNN-DATA and PINS were matched in training data, neural architecture, and
	computational budget. Both use a
	$21\!\to\!96\!\to\!96\!\to\!96\!\to\!7$ architecture and were trained on
	2{,}100 design--scenario pairs, with 450 design-disjoint validation pairs and
	450 frozen test pairs. SNN-DATA minimizes the normalized data loss alone,
	whereas PINS augments it with the two-diode and voltage-ordering residuals.
	Validation-only tuning selected $\lambda_{\mathrm{phys}}=0.3$.
	The predictive comparison is intentionally non-decisive. Validation NRMSE is
	0.301721 for SNN-DATA and 0.304250 for PINS; on the frozen test set the
	corresponding values are 0.350842 and 0.353172. The difference is therefore
	less than 1\%, with neither model exhibiting a meaningful aggregate accuracy
	advantage. Target-level performance is similarly mixed
	(Fig.~\ref{fig:current_parity}). This near tie is important: deployment
	cannot be justified retrospectively by conventional held-out error alone.
	
\noindent	The degradation experiments provide complementary evidence. Across four
	label-noise levels and four reduced-data budgets, PINS obtained the lower
	mean test NRMSE in six of eight conditions. The advantage is modest rather
	than universal, but it is consistent with a robustness benefit from
	physics-informed regularization when training information is degraded.
	
	%---------------------------------------------------------------------
	\subsection{High-Fidelity Operational Selection of the Surrogate}
	\label{sec:results_operational_selection}
	%---------------------------------------------------------------------
	
\noindent	Held-out error measures performance under the sampled data distribution, but
	a multiobjective optimizer deliberately concentrates queries near the current
	Pareto set and active constraint boundaries. The decisive comparison was
	therefore conducted under this optimizer-induced query distribution. Each
	surrogate independently drove matched NSGA-II searches over the same declared
	design domain, after which the resulting candidate sets were re-evaluated by
	the HF simulator on one common fixed
	$N_{\mathrm{scr}}=100$ scenario batch. This stage is strictly \emph{operational screening}; it is not the independent
	500-scenario reliability certification reported later.	The difference is pronounced despite the near-identical frozen-test errors
	(Table~\ref{tab:current_operational}). SNN-DATA achieved only 44.4\%
	operational HF agreement, below the pre-specified 60\% decision-validity
	floor, whereas PINS achieved 93.4\%. Both models exhibited zero false
	feasibility, but false infeasibility was 55.6\% for SNN-DATA and only 6.6\%
	for PINS. Thus, the principal weakness of the data-only model was not unsafe
	acceptance of infeasible designs, but rejection of a large fraction of
	HF-feasible candidates.
	
	\begin{table}
		\centering
		\caption{Independent HF operational screening under the common
			100-scenario batch.}
		\label{tab:current_operational}
		\scriptsize
		\setlength{\tabcolsep}{3pt}
		\renewcommand{\arraystretch}{1.03}
		\resizebox{\columnwidth}{!}{%
			\begin{tabular}{lccccc}
				\toprule
				\textbf{Model} &
				\textbf{$n$} &
				\textbf{HF agree.} &
				\textbf{FF} &
				\textbf{FI} &
				\textbf{Dominated}\\
				\midrule
				SNN-DATA & 72  & 44.4\% & 0.0\% & 55.6\% & 72/72 (100\%)\\
				PINS     & 183 & 93.4\% & 0.0\% & 6.6\%  & 0/183 (0\%)\\
				\bottomrule
		\end{tabular}}
		\vspace{1mm}
		
		\footnotesize
		FF: false feasible; FI: false infeasible.
		Decision-validity threshold: 60\%.
	\end{table}
	
\begin{figure}
		\centering
		\includegraphics[width=\linewidth]{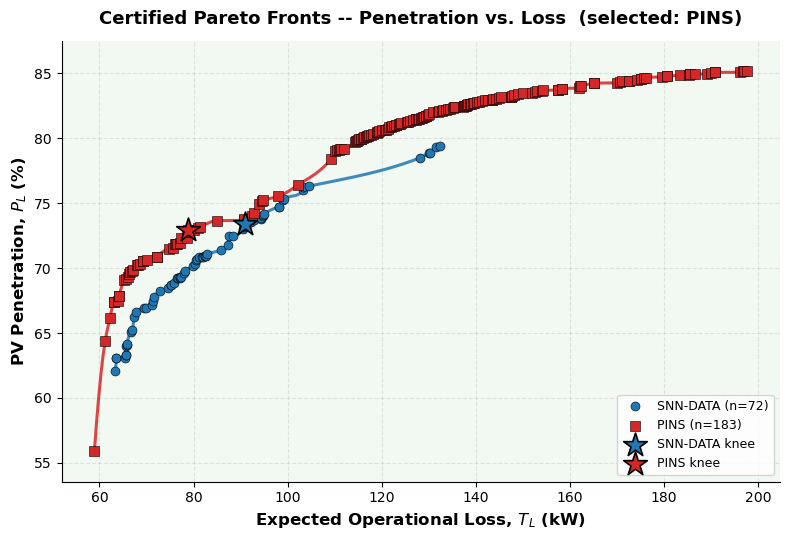}
		\caption{Search-stage Pareto sets under the common HF screening batch.
			Every SNN-DATA candidate is dominated by at least one PINS candidate,
			whereas no PINS candidate is dominated by an SNN-DATA design.}
		\label{fig:current_pareto}
	\end{figure}

	The objective-space evidence reinforces the operational result. Every one of
	the 72 SNN-DATA search-stage designs is dominated by at least one PINS
	candidate, while none of the 183 PINS candidates is dominated by an
	SNN-DATA design (Fig.~\ref{fig:current_pareto}). This is a direct geometric
	property of the evaluated candidate sets, not a statistical inference.
	A supplementary Mann--Whitney analysis also finds different marginal loss and
	penetration distributions ($p<0.0001$), but the operational conclusion rests
	primarily on common-batch HF agreement, false-feasibility/false-infeasibility,
	and cross-dominance.
	
\noindent	PINS was therefore selected for downstream optimization because it preserves
	decision-relevant HF structure much more effectively in the optimizer-induced
	region, not because it minimizes conventional test error.
		The computational gain is substantial. 
		On the executed computational platform, the 60-scenario HF evaluation
		required 286.0~ms per design, compared with 0.46~ms per
		design-equivalent for vectorized PINS inference, corresponding to an
		approximately $620\times$ search-stage speedup.
		 This reduction is what makes repeated stochastic
	multiobjective evaluation practical. Operational superiority should nevertheless not be confused with uniformly
	accurate distributional prediction. The supplementary HF quantile audit shows
	that PINS underpredicts the HF 95th-percentile $T_L$ by approximately 30\%
	(Fig.~\ref{fig:current_tail_fidelity}, Appendix~\ref{appdiagnostics}).
	This residual upper-tail error strengthens the case for recomputing all final
	objectives and reliability statistics using the HF simulator.
	
	%---------------------------------------------------------------------
	\subsection{Periodic Decision-Gated High-Fidelity Enrichment}
	\label{sec:results_enrichment}
	%---------------------------------------------------------------------
	
	\noindent Periodic enrichment returns optimizer-relevant designs from Pareto-critical,
	constraint-boundary, and surrogate-disagreement regions to the HF model.
	In the reported audited update, 117 unique designs generated
	5{,}616 valid HF-labeled design--scenario pairs
	(Table~\ref{tab:enrichconfig}, Appendix~\ref{appenrich}). 	On the protected audit set, the candidate enrichment update reduced mean
	normalized prediction error from 0.1829 to 0.1113, corresponding to a
	\textbf{39.1\% reduction}. This demonstrates that optimizer-targeted HF
	sampling can substantially reduce local prediction error in regions
	emphasized by the evolving search.
	
	%---------------------------------------------------------------------
	\subsection{Independent HF Certification and Final Pareto Decision}
	\label{sec:results_final_optimization}
	\label{sec:results_certification}
	\label{sec:topsis}
	%---------------------------------------------------------------------
	
\noindent	Final planning claims are based exclusively on independent HF evaluation.
	The search-stage population was reduced to 80 candidate designs, each of
	which was subsequently recomputed on a disjoint
	$N_{\mathrm{cert}}=500$ scenario set. Reliability was quantified using the
	exact one-sided 95\% Clopper--Pearson lower confidence bound rather than the
	raw observed safe fraction.
	
	All 80 candidates were HF-valid, penetration-compliant, loss-compliant, and
	satisfied the required 0.90 reliability threshold. Consequently,
	\textbf{80/80 candidates were independently HF-certified}. Non-dominated
	filtering retained 46 certified Pareto designs
	(Table~\ref{tab:final_certification}).
	
	\begin{table}
		\centering
		\caption{Independent 500-scenario HF certification.}
		\label{tab:final_certification}
		\scriptsize
		\setlength{\tabcolsep}{3pt}
		\renewcommand{\arraystretch}{0.95}
		\begin{tabular}{lr}
			\toprule
			\textbf{Quantity} & \textbf{Result}\\
			\midrule
			Candidates entering certification & 80\\
			HF-valid & 80/80\\
			CP95 safety-certified & 80/80\\
			Penetration-compliant & 80/80\\
			Loss-compliant & 80/80\\
			\textbf{Fully HF-certified} & \textbf{80/80 (100\%)}\\
			HF non-dominated designs & \textbf{46}\\
			Pareto $E[T_L]$ range & 264.49--300.64~kW\\
			Pareto $E[\Pi]$ range & 53.15--71.58\%\\
			TOPSIS $E[T_L]$ & \textbf{297.92~kW}\\
			TOPSIS $E[\Pi]$ & \textbf{70.74\%}\\
			Empirical $p_{\mathrm{safe}}$ & 0.982\\
			CP95 lower bound & \textbf{0.9688}\\
			Required probability & 0.90\\
			TOPSIS closeness score & 0.6922\\
			\bottomrule
		\end{tabular}
	\end{table}
	
\begin{figure}
		\centering
		\includegraphics[width=\linewidth]{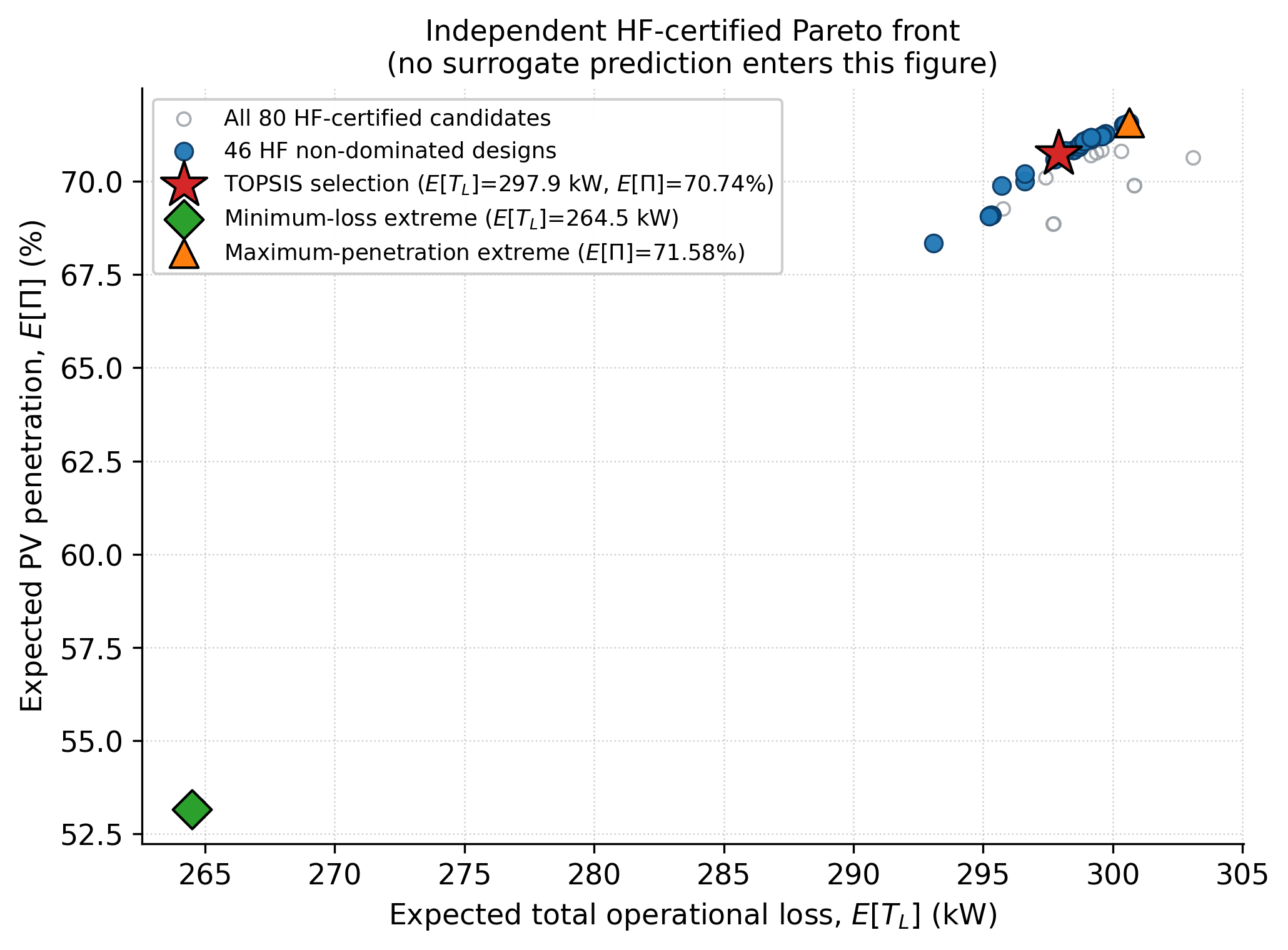}
		\caption{Independent 500-scenario HF-certified Pareto front. Filled
			markers denote the 46 non-dominated designs; the TOPSIS compromise is
			identified after certification. All displayed objective values are
			recomputed using the HF simulator.}
		\label{fig:current_hf_pareto}
	\end{figure}

	The independently evaluated Pareto front spans
	$E[T_L]=264.49$--$300.64$~kW and
	$E[\Pi]=53.15$--$71.58\%$, demonstrating a genuine loss--penetration
	trade-off rather than a unique deterministic hosting limit.
	Figure~\ref{fig:current_hf_pareto} is therefore the principal planning result
	of the study: every plotted non-dominated point is obtained from the
	independent HF certification stage rather than from surrogate prediction.
	
	TOPSIS \cite{hwang1981topsis} is deliberately applied only after HF
	certification and therefore has no role in determining feasibility. Under
	equal loss--penetration weights, the selected compromise has
	\[
	E[T_L]=297.92~\mathrm{kW},
	\qquad
	E[\Pi]=70.74\%,
	\]
	with empirical safety probability $p_{\mathrm{safe}}=0.982$ and a
	one-sided 95\% Clopper--Pearson lower bound
	$p^{L}_{\mathrm{CP},0.95}=0.9688$, comfortably above the required 0.90.
	Every other member of the certified Pareto front remains an admissible
	planning alternative; TOPSIS simply maps an explicit preference onto an
	already feasible and non-dominated set.
	
	Preference sensitivity confirms this interpretation. Seven tested
	loss--penetration weight pairs select five distinct certified designs,
	spanning approximately 53.2--71.6\% expected penetration. Hence, the
	certified Pareto front, rather than any one TOPSIS point, is the fundamental
	decision product.
	
	\begin{table}
		\centering
		\caption{Representative TOPSIS weight sensitivity on the certified
			Pareto set.}
		\label{tab:current_weights}
		\scriptsize
		\setlength{\tabcolsep}{4pt}
		\renewcommand{\arraystretch}{0.95}
		\begin{tabular}{rrrr}
			\toprule
			$w_L/w_\Pi$ & $E[T_L]$ [kW] & $E[\Pi]$ [\%] & Score\\
			\midrule
			0.2/0.8 & 300.4 & 71.51 & 0.898\\
			\textbf{0.5/0.5} &
			\textbf{297.9} &
			\textbf{70.74} &
			\textbf{0.692}\\
			0.8/0.2 & 264.5 & 53.15 & 0.646\\
			\bottomrule
		\end{tabular}
	\end{table}
	
	%---------------------------------------------------------------------
	\subsection{Representative Operational-Loss Decomposition}
	\label{sec:results_loss_decomposition}
	%---------------------------------------------------------------------
	
\noindent	After establishing the independently certified decision space, the physical
	origin of the loss objective can be examined without conflating
	component-level accounting with stochastic optimization outcomes.
	Table~\ref{tab:loss_decomposition} decomposes $T_L$ at one representative
	HF operating condition with array MPP of 1{,}301.3~kW and instantaneous
	penetration of 27.1\%. This diagnostic operating point is intentionally
	distinct from the TOPSIS stochastic expectation
	$E[T_L]=297.92$~kW.
	\begin{table}
		\centering
		\caption{Loss decomposition for a representative HF operating condition.}
		\label{tab:loss_decomposition}
		\scriptsize
		\setlength{\tabcolsep}{3.5pt}
		\renewcommand{\arraystretch}{0.95}
		
		\begin{tabular}{@{}lrr@{}}
			\toprule
			\textbf{Component} & \textbf{Loss [kW]} & \textbf{Share [\%]} \\
			\midrule
			DC cable            & 0.16   & 0.1 \\
			DC conditioning     & 6.51   & 4.2 \\
			Current curtailment & 0.00   & 0.0 \\
			Inverter conversion & 23.30  & 15.1 \\
			AC conditioning     & 12.71  & 8.2 \\
			Power curtailment   & 0.00   & 0.0 \\
			AC cable            & 0.13   & 0.1 \\
			\midrule
			\textbf{PV-side subtotal} & \textbf{42.82} & \textbf{27.7} \\
			Feeder network      & 111.57 & 72.3 \\
			\midrule
			\textbf{Total}      & \textbf{154.39} & \textbf{100.0} \\
			\bottomrule
		\end{tabular}
	\end{table}
	
\noindent	The energy balance closes to approximately
	$1.9\times10^{-13}$~kW. Two features are particularly important.
	First, inverter conversion is the dominant PV-side contribution,
	23.30~kW or 15.1\% of total operational loss, whereas the individual cable
	terms are negligible at this operating point. Second, feeder-network loss
	contributes 111.57~kW, or 72.3\% of $T_L$, exceeding all PV-side loss
	mechanisms combined. This asymmetry explains why electrical siting can dominate local
	component-sizing changes. Improving cable dimensions cannot reproduce the
	network-level effect of moving a large injection electrically closer to, or
	farther from, heavily loaded feeder sections. The decomposition therefore
	provides a physical interpretation of the strong PCC sensitivity reported in
	Section~\ref{sec:results_sensitivity}.
	
	%---------------------------------------------------------------------
	\subsection{Multi-Feeder Engineering Validation}
	\label{sec:results_multifeeder}
	%---------------------------------------------------------------------
	
\begin{figure*}
		\centering
		\includegraphics[width=1.0\linewidth,height=12.0cm]{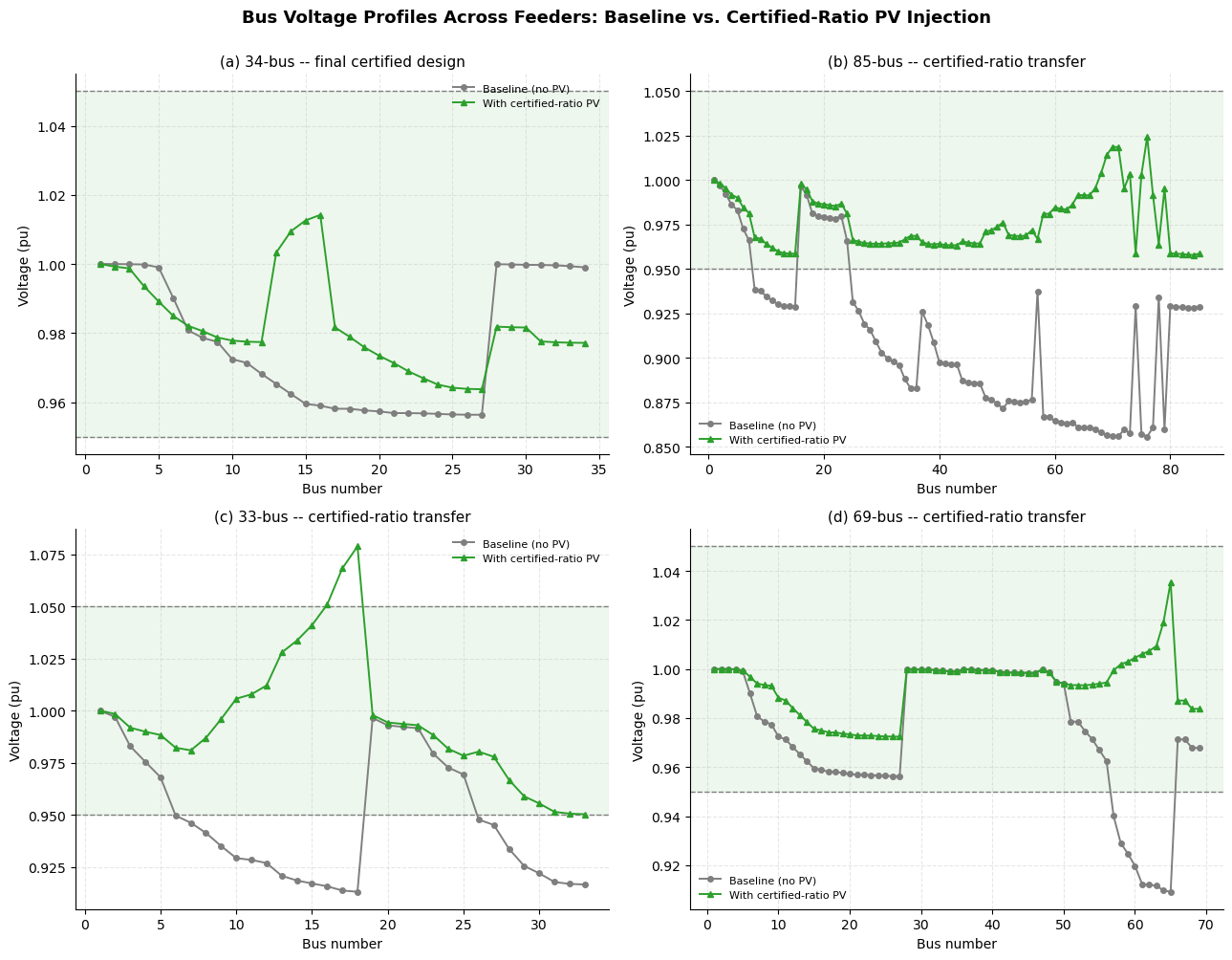}
		\caption{Bus voltage profiles, baseline versus PV injection. Panel (a)
			shows the primary 34-bus TOPSIS design under one representative scenario
			(instantaneous penetration 107.25\%); panels (b)--(d) impose the certified
			expected penetration $E[\Pi]=70.74\%$ on the 85-, 33-, and 69-bus
			feeders. Panel (c) exposes the 33-bus overvoltage above 1.05~pu.}
		\label{fig:current_voltage_multi}
	\end{figure*}

\begin{figure}
		\centering
		\includegraphics[width=\linewidth]{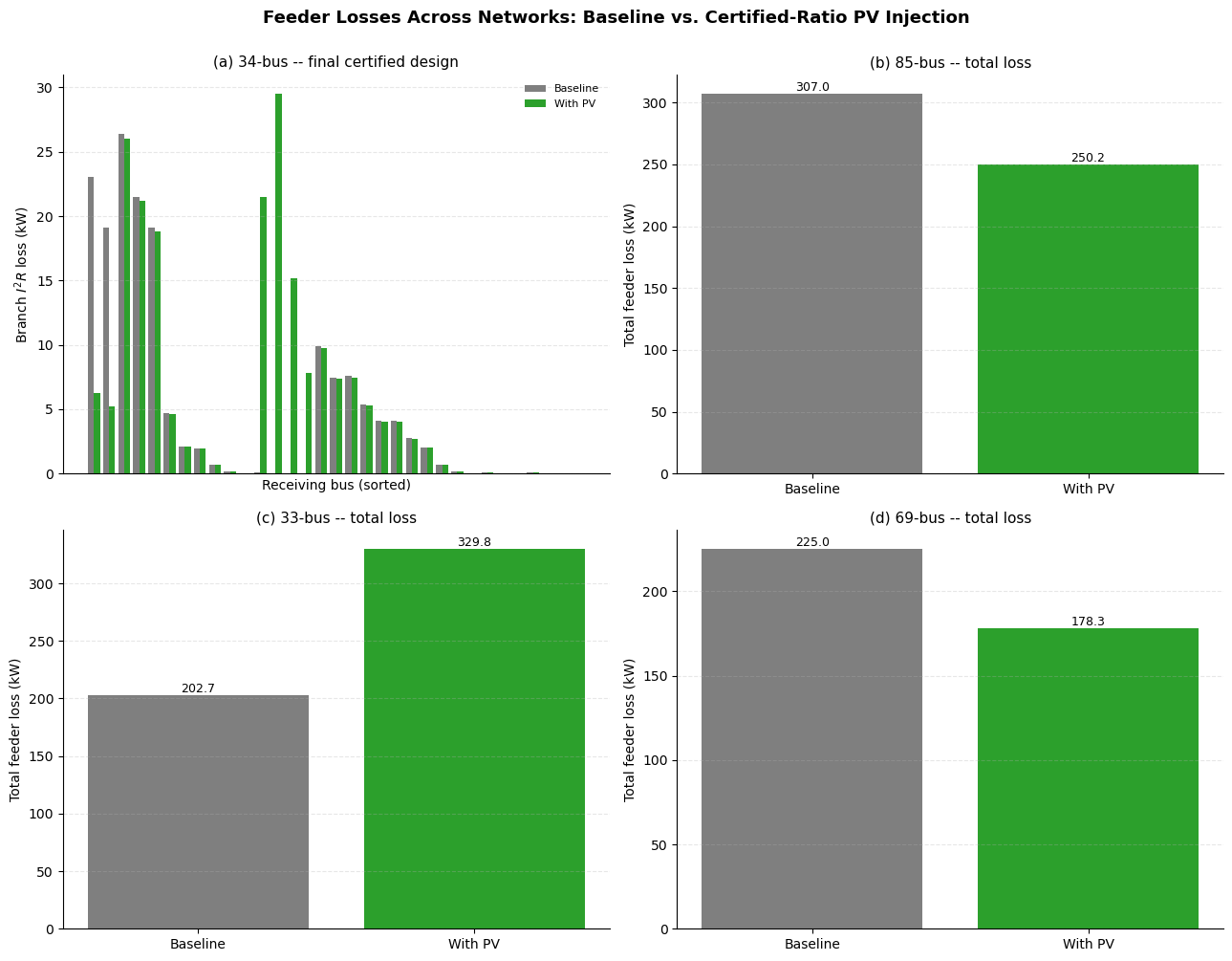}
		\caption{Feeder-network resistive loss $T_{\mathrm{network}}$, baseline
			versus PV injection. Network loss decreases on the 85- and 69-bus
			feeders but increases on the 34- and 33-bus systems, demonstrating that
			PV loss response is topology- and siting-dependent.}
		\label{fig:current_losses_multi}
	\end{figure}

\noindent	The final engineering stress test asks whether the certified penetration level
	transfers unchanged to networks that played no role in optimization or
	certification. It is therefore a test of \emph{methodological portability},
	not a new reliability certificate.	The distinction between the design feeder and transfer feeders is important.
	The primary 34-bus panel represents the TOPSIS design under one specific
	operating scenario, for which instantaneous penetration is 107.25\%.
	The 85-, 33-, and 69-bus stress tests instead impose the independently
	certified expected penetration ratio
	$E[\Pi]=70.74\%$. These quantities are not contradictory: one is a
	scenario-level ratio and the other a stochastic expectation. The results are strongly topology-dependent
	(Figs.~\ref{fig:current_voltage_multi}--\ref{fig:current_losses_multi};
	Table~\ref{tab:multi_feeder}). On the 85-bus feeder,
	$V_{\min}$ improves from 0.8557 to 0.9579~pu, all 69 baseline violations
	are eliminated, and $T_{\mathrm{network}}$ decreases by 18.5\%.
	The 69-bus feeder similarly clears all nine baseline voltage violations,
	with $V_{\min}$ increasing from 0.9092 to 0.9724~pu and network loss
	decreasing by 20.7\%. The 33-bus feeder exhibits the opposite limitation. Its minimum voltage improves from 0.9131 to 0.9503~pu and the violation count decreases from
	21 to 3, yet the injection region reaches approximately 1.081~pu, exceeding
	the 1.05~pu upper limit, while network loss increases from 202.7 to
	329.8~kW, a 62.7\% increase. Under the representative 34-bus operating
	condition, $T_{\mathrm{network}}$ also increases by 25.3\%, although voltage
	limits remain satisfied.
	
	\begin{table}
		\centering
		\caption{Cross-feeder transfer performance. Loss values denote
			$T_{\mathrm{network}}$, not the full eight-term $T_L$.}
		\label{tab:multi_feeder}
		\scriptsize
		\setlength{\tabcolsep}{2.4pt}
		\renewcommand{\arraystretch}{1.03}
		\resizebox{\columnwidth}{!}{%
			\begin{tabular}{lccccccc}
				\toprule
				\textbf{Feeder} &
				\textbf{$V_{\min}^{0}$} &
				\textbf{$V_{\min}^{PV}$} &
				\textbf{Viol.$^{0}$} &
				\textbf{Viol.$^{PV}$} &
				\textbf{$T_{\rm net}^{0}$} &
				\textbf{$T_{\rm net}^{PV}$} &
				\textbf{$\Delta T_{\rm net}$}\\
				\midrule
				34-bus & 0.9563 & 0.9637 & 0  & 0 & 163.5 & 204.9 & $+25.3\%$\\
				85-bus & 0.8557 & 0.9579 & 69 & 0 & 307.0 & 250.2 & $-18.5\%$\\
				33-bus & 0.9131 & 0.9503 & 21 & 3 & 202.7 & 329.8 & $+62.7\%$\\
				69-bus & 0.9092 & 0.9724 & 9  & 0 & 225.0 & 178.3 & $-20.7\%$\\
				\bottomrule
		\end{tabular}}
		\vspace{1mm}
		
		\footnotesize
		Voltages are in pu; violations use the 0.95--1.05~pu band.
	\end{table}
	
\noindent	The mixed response is one of the most informative engineering results of the
	study. Distributed PV does not produce a universal loss reduction simply by
	increasing local generation. Depending on network impedance and injection
	location, PV can reduce downstream current and correct undervoltage, or
	produce reverse power flow, larger branch-current magnitudes, and local
	overvoltage. The 33-bus case therefore demonstrates directly why a penetration
	ratio certified on one feeder must be re-verified, and where necessary
	re-optimized for PCC location, before being applied to another.
	
	%------------
	
	\subsection{Robustness and Secondary Planning Checks}
	\label{sec:results_secondary}
	\label{sec:results_sensitivity}
	\label{sec:results_algorithms}
	\label{sec:results_budget}
	\label{sec:results_capacity}
	\label{sec:results_threshold}
	
	\noindent
	Secondary analyses confirm that the main conclusions are not driven by a
	single training condition, optimizer, design perturbation, or planning bound.
	PINS outperformed SNN-DATA in six of eight noise/data-budget cases, indicating
	moderate robustness rather than universal superiority. HF sensitivity analysis
	identified PCC location as the dominant local design variable, with re-siting
	changing loss by up to approximately 26\%, while most other perturbations had
	much smaller effects. Across NSGA-II, NSGA-III, SPEA2, and SMS-EMOA, normalized
	hypervolume differences were not significant (Friedman $p=0.392$); MOEA/D did
	not complete any of the eight runs under the implemented constraint handling.
	Detailed sensitivity and algorithm results are given in
	Appendix~\ref{appdiagnostics}.The HF-budget study showed a direct link between training information and
	planning outcome: increasing the budget from 50 to 1{,}000 evaluations reduced
	held-out NRMSE from 0.930 to 0.463, with certified designs increasing from none
	at 50 evaluations to 35 at 100 and 47 at 1{,}000; the corresponding TOPSIS
	penetration increased from 30.3\% to 63.2\%. Penetration is defined throughout
	using scenario-dependent feeder demand, not historical peak load. Finally,
	sweeping the penetration ceiling from 70\% to 200\% produced no binding upper
	limit, indicating that the reported solutions are governed by network, loss,
	and reliability constraints rather than by the imposed search ceiling.\\

	\noindent
	The results establish a clear separation between optimization, learning, and
	engineering certification. Mathematically, the solution is a
	reliability-qualified approximation to a chance-constrained Pareto set, with
	SAA supporting stochastic search, independent 500-scenario HF evaluation
	establishing reliability, and TOPSIS applied only after feasibility and
	non-dominance are verified. From the AI perspective, predictive accuracy alone
	is insufficient: SNN-DATA and PINS differ by less than 1\% in frozen-test
	NRMSE, yet their HF decision agreement is 44.4\% and 93.4\%, respectively,
	while targeted HF enrichment reduces audited local prediction error by 39.1\%.
	Residual upper-tail loss error of approximately 30\% further justifies retaining
	HF evaluation for final decisions. Engineering evidence is consistent with this
	hierarchy: feeder-network loss contributes 72.3\% of representative total loss,
	PCC location is the dominant local sensitivity, and cross-feeder tests show
	strong topology dependence. Thus, the surrogate provides computational
	acceleration, whereas HF physics and independent statistical certification
	determine final feasibility, reliability, and transferability.

	%=====================================================================
	\section{Conclusion}
	\label{sec:conclusion}
	%=====================================================================
\noindent	This study developed a decision-focused surrogate-assisted framework for
	stochastic PV penetration planning in which a calibrated two-diode PV model
	and nonlinear AC power flow remain the final engineering authority while
	machine learning accelerates repeated stochastic multiobjective evaluation.
	Although SNN-DATA and PINS achieved nearly identical frozen-test performance
	(NRMSE 0.3508 and 0.3532), optimizer-realistic HF screening separated their
	decision performance sharply: operational agreement was 44.4\% for
	SNN-DATA and 93.4\% for PINS, and every one of the 72 SNN-DATA search-stage
	designs was dominated by at least one PINS design. PINS was therefore
	selected on decision-level evidence rather than average predictive accuracy.
	At the $N_{\mathrm{gen}}=60$ scenario budget, surrogate inference reduced
	evaluation time from 286.0 to 0.46~ms per design-equivalent
	($\approx620\times$), while periodic targeted HF enrichment reduced audited
	mean normalized prediction error by 39.1\%. Independent 500-scenario HF evaluation subsequently certified all 80 screened
	candidates at the required 0.90 reliability level and retained 46
	non-dominated designs. The equal-weight TOPSIS compromise achieves
	$E[\Pi]=70.74\%$ at $E[T_L]=297.92$~kW, with empirical safety probability
	0.982 and a one-sided 95\% Clopper--Pearson lower bound of 0.9688. The
	certified front spans approximately 53.15--71.58\% expected penetration,
	showing that the primary planning result is a reliability-qualified
	loss--penetration trade-off rather than a single unconditioned capacity
	number. Cross-feeder transfer provides the principal engineering qualification to
	that result. The transferred penetration level removes severe baseline
	voltage violations and reduces network losses on the 85- and 69-bus feeders,
	whereas the 33-bus system develops local overvoltage and a 62.7\% increase
	in network loss. Together with the 72.3\% network contribution to
	representative operational loss and the dominant sensitivity to PCC location,
	this demonstrates that hosting capability is intrinsically topology- and
	siting-dependent. The resulting framework therefore assigns distinct roles
	to its components: surrogates provide speed, physics-informed structure
	supports decision robustness, HF feedback improves local fidelity, and
	independent HF simulation provides final statistical and engineering
	authority. A certified penetration ratio should consequently be interpreted
	as a network-specific planning result and a starting point for feeder-specific
	siting and capacity assessment, not as a universally portable certificate.

	\begin{figure*}
	\centering
	\includegraphics[
	width=0.80\textwidth,
	keepaspectratio
	]{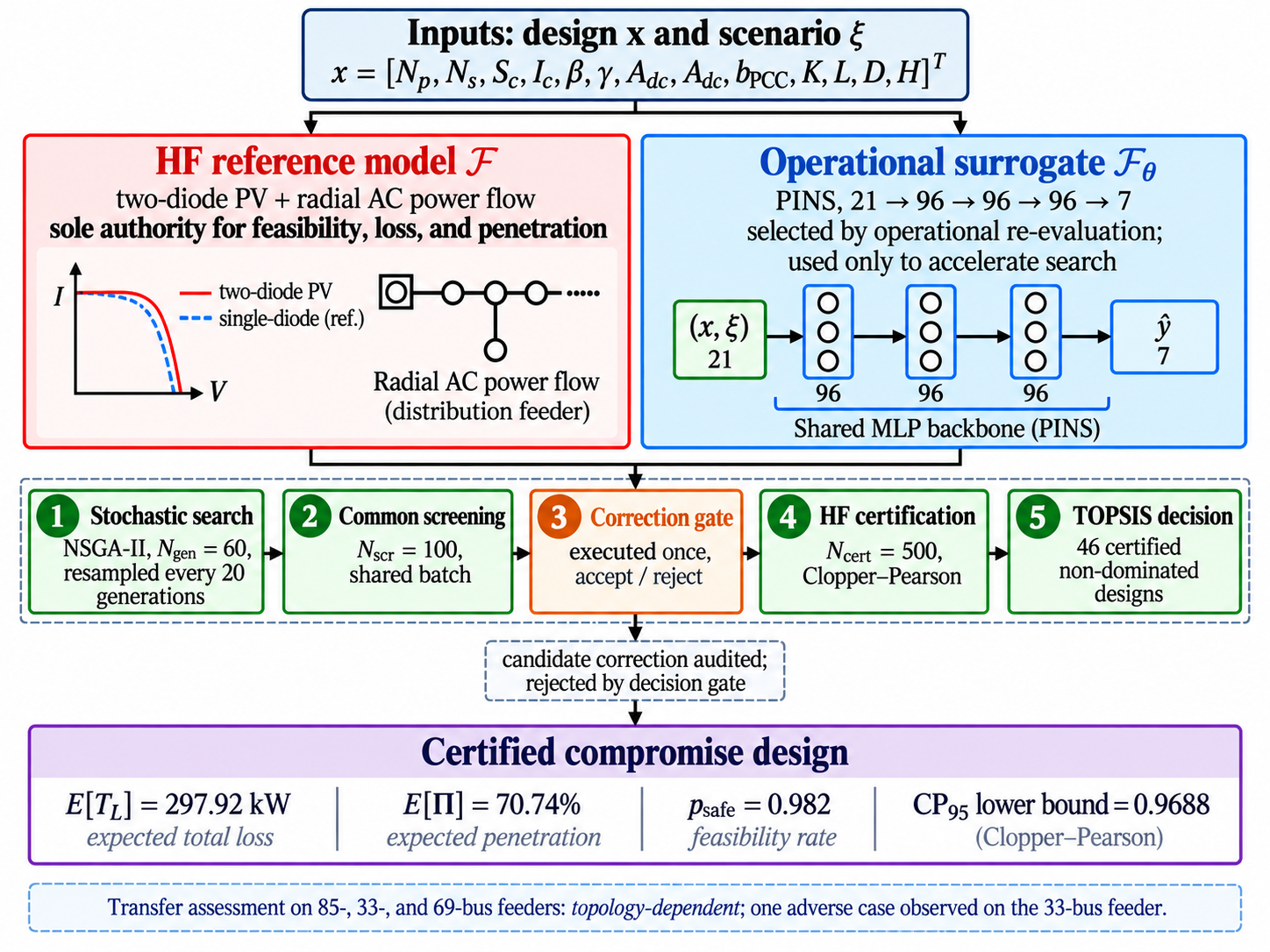}
	\caption{Decision-gated surrogate-assisted planning framework integrating
		HF evaluation, surrogate optimization, targeted enrichment, independent
		certification, and cross-feeder assessment.}
	\label{fig:current_framework_overview}
\end{figure*}

	\section*{Acknowledgment}
\noindent	The authors gratefully acknowledge the Michigan Institute for Data and AI in Society (MIDAS), University of Michigan, through the Schmidt AI in Science African
Faculty Fellowship MIDAS at the University of Michigan for funding and supporting the interdisciplinary research environment that enabled this work, and the University of Michigan--Dearborn for institutional support.
	
	\section*{Declaration of Generative AI and AI-Assisted Technologies}
\noindent	During manuscript preparation, the authors used ChatGPT (OpenAI) and Claude
	(Anthropic) for language editing, manuscript organization, \LaTeX\ formatting,
	and coding assistance. All equations, numerical results, analyses, scientific
	interpretations, and conclusions were independently reviewed and verified by
	the authors, who take full responsibility for the content of the manuscript.
	
	%=====================================================================
	%  Elsevier-style appendix numbering (Appendix A, A.1, (A.1), Table A.1, Fig. A.1)
	%=====================================================================
	\renewcommand{\thesection}{Appendix~\Alph{section}}
	\renewcommand{\thesubsection}{\Alph{section}.\arabic{subsection}}
	\renewcommand{\theequation}{\Alph{section}.\arabic{equation}}
	\renewcommand{\thetable}{\Alph{section}.\arabic{table}}
	\renewcommand{\thefigure}{\Alph{section}.\arabic{figure}}
	\setcounter{section}{0}
	%=====================================================================
	\section{Supporting Physical, Statistical, and Reproducibility Details}
	\label{app:supplementary}
	\setcounter{equation}{0}
	\setcounter{table}{0}
	\setcounter{figure}{0}
	%=====================================================================
	
	\noindent This appendix records supporting physical definitions, stochastic-sampling
	assumptions, enrichment settings, and diagnostic evidence needed to interpret
	or reproduce the main study. It is intentionally limited to material that
	supports, but does not duplicate, the principal results.
	
	%---------------------------------------------------------------------
	\subsection{Notation and Reliability Terminology}
	\label{appterm}
	%---------------------------------------------------------------------
	
	HF denotes the two-diode-plus-AC-power-flow reference simulator; PINS and
	SNN-DATA denote the matched physics-penalized and data-only neural surrogates;
	SAA denotes sample-average approximation; PCC denotes point of common
	coupling; and CP95 denotes the one-sided 95\% Clopper--Pearson lower
	confidence bound.
	
	PV penetration is defined throughout as
	\begin{equation}
		\Pi(\mathbf{x})
		=
		\mathbb E_\xi
		\left[
		100\frac{P_{\mathrm{PCC}}(\mathbf{x},\xi)}
		{P_D(\xi)}
		\right],
		\label{eq:penetration_def}
	\end{equation}
	and should therefore not be interpreted as a ratio against fixed historical
	peak demand or installed feeder capacity.
	
	The terms \emph{search-time}, \emph{screened}, and \emph{certified} also have
	different meanings. Search-time statistics use $N_{\mathrm{gen}}=60$;
	comparative operational screening uses the common
	$N_{\mathrm{scr}}=100$ batch; and only results recomputed on the independent
	$N_{\mathrm{cert}}=500$ HF set are described as certified.
	
	%---------------------------------------------------------------------
	\subsection{Two-Diode Residual Used by the Physics-Penalized Surrogate}
	\label{apptd}
	%---------------------------------------------------------------------
	
	At cell level, the HF PV model satisfies
	\begin{equation}
		\begin{aligned}
			I_{\mathrm{cell}}
			={}& I_{\mathrm{ph}}
			-I_{01}\left[
			\exp\!\left(
			\frac{V_{\mathrm{cell}}+I_{\mathrm{cell}}R_s}{n_1V_t}
			\right)-1
			\right]
			\\
			&-I_{02}\left[
			\exp\!\left(
			\frac{V_{\mathrm{cell}}+I_{\mathrm{cell}}R_s}{n_2V_t}
			\right)-1
			\right]
			-\frac{V_{\mathrm{cell}}+I_{\mathrm{cell}}R_s}{R_{\mathrm{sh}}},
		\end{aligned}
		\label{eq:two_diode_cell_v3}
	\end{equation}
	where
	\[
	V_t=\frac{kT_c}{q},\qquad
	I_{\mathrm{ph}}
	=
	\left[
	I_{\mathrm{sc}}
	+\mu_{I,\mathrm{sc}}(T_c-T_{\mathrm{ref}})
	\right]
	\frac{G_T}{G_{\mathrm{ref}}}.
	\]
	
	Because the neural network outputs array-level voltage and current, its
	predictions are transformed using
	\begin{equation}
		\widehat V_m
		=
		\frac{\widehat V_{\mathrm{pv}}}{N_sN_{\mathrm{cell}}},
		\qquad
		\widehat I_m
		=
		\frac{\widehat I_{\mathrm{pv}}}{N_p},
		\label{eq:surrogate_unpack_v3}
	\end{equation}
	before the two-diode residual is evaluated. The implemented scaled residual is
	\begin{equation}
		\widetilde{\mathcal R}_I
		=
		\frac{\mathcal R_I}{s_I+\epsilon},
		\label{eq:residual_scaled}
	\end{equation}
	and the PV contribution to the PINS loss is
	\begin{equation}
		\mathcal L_{\mathrm{PV}}
		=
		\frac{1}{N_d}
		\sum_{k=1}^{N_d}
		\left[
		\operatorname{asinh}
		\left(
		\widetilde{\mathcal R}_I^{(k)}
		\right)
		\right]^2.
		\label{eq:pins_pv_loss}
	\end{equation}
	Together with the voltage-ordering term of Section~\ref{sec:pins_v2}, this
	penalizes physically inconsistent predictions without requiring the surrogate
	to reproduce the complete AC bus-state vector.
	
	%---------------------------------------------------------------------
	\subsection{Interconnection and Network-Loss Accounting}
	\label{appinterconnection}
	%---------------------------------------------------------------------
	
	The inverter-current limiter is
	\[
	\kappa_c
	=
	\min\!\left(1,\frac{I_c}{I_{\mathrm{pv}}}\right),
	\]
	whereas the inverter apparent-power rating $S_c$ is checked separately after
	conversion. Thus satisfying the DC-current limiter does not imply satisfaction
	of the AC rating.
	
	Ohmic conductor loss follows
	\[
	L=I^2R,\qquad
	R=\frac{\rho\ell}{A}.
	\]
	The DC cable, dedicated AC interconnection cable, and feeder-network branches
	are therefore evaluated as distinct physical elements. This separation
	prevents the same resistance or dissipation mechanism from entering
	$T_L$ more than once.
	
	%---------------------------------------------------------------------
	\subsection{Scenario Independence and Clopper--Pearson Interpretation}
	\label{appscenario}
	%---------------------------------------------------------------------
	
	A stochastic scenario consists of one complete historical weather state and
	one independently sampled demand multiplier,
	\[
	m_L\sim\operatorname{Lognormal}(\mu_L,\sigma_L^2),
	\qquad
	\sigma_L=0.10,
	\]
	truncated to $[0.70,1.30]$. Weather variables within a scenario are therefore
	dependent exactly as they appear in the sampled historical row, whereas
	different scenario draws are generated independently with replacement.
	
	The final safety indicators
	\[
	\mathbb 1[\mathcal S(\mathbf{x},\xi_\nu)]
	\]
	are consequently treated as independent Bernoulli observations for the
	Clopper--Pearson calculation. The statistical independence assumption applies
	between Monte Carlo scenario draws; it does not require irradiance,
	temperature, solar geometry, and albedo inside a single scenario to be
	independent.
	
	%---------------------------------------------------------------------
	\subsection{Periodic HF-Enrichment Configuration}
	\label{appenrich}
	%---------------------------------------------------------------------
	
	At each scheduled enrichment audit, the current optimization population is
	sampled from Pareto-critical, constraint-boundary, and
	surrogate-disagreement regions. Table~\ref{tab:enrichconfig} gives the
	configuration of the representative audited round reported in
	Section~\ref{sec:results_enrichment}.
	
	\begin{table}
		\centering
		\caption{Representative periodic HF-enrichment configuration.}
		\label{tab:enrichconfig}
		\scriptsize
		\setlength{\tabcolsep}{4pt}
		\begin{tabular}{lcc}
			\toprule
			\textbf{Acquisition} & \textbf{Requested} & \textbf{Purpose}\\
			\midrule
			Pareto & 30 & Objective-critical\\
			Boundary & 45 & Constraint-critical\\
			Disagreement & 45 & Model-disagreement\\
			\midrule
			Unique designs & 117 & After deduplication\\
			Valid HF pairs & 5{,}616 & HF-labeled samples\\
			\bottomrule
		\end{tabular}
	\end{table}
	
	The numerical quotas describe the audited round, not a claim that every
	periodic audit produces exactly 117 unique points: overlap among acquisition
	sets may vary with the current optimization population. Candidate surrogate
	updates are accepted only when the prediction-error, false-feasibility, and
	decision-agreement gates all pass simultaneously.
	
	%---------------------------------------------------------------------
	\subsection{Supplementary Physical, Robustness, and Sensitivity Diagnostics}
	\label{appdiagnostics}
	%---------------------------------------------------------------------
	
	\begin{figure}
		\centering
		\includegraphics[width=\linewidth]{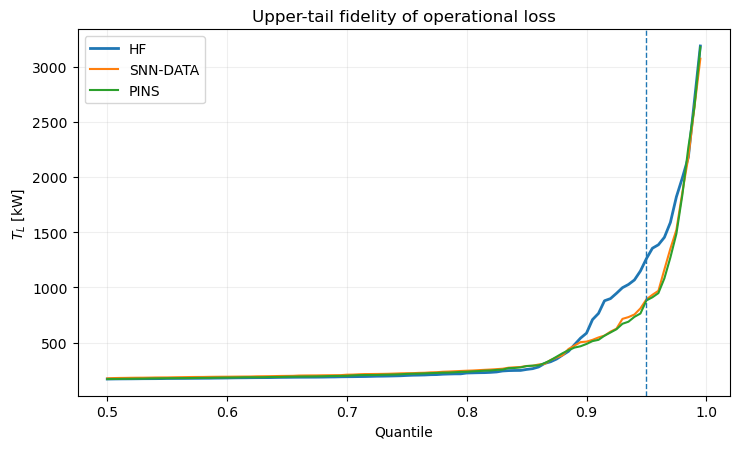}
		\caption{HF and surrogate $T_L$ quantiles over the 0.50--0.99 probability
			range.}
		\label{fig:current_tail_fidelity}
	\end{figure}
	
	The tail comparison in Fig.~\ref{fig:current_tail_fidelity} is intentionally
	retained because it identifies an important limitation of the selected
	surrogate. PINS underpredicts the HF $T_L$ 95th percentile by approximately
	30\%. Consequently, its superior operational feasibility agreement should not
	be interpreted as uniformly accurate tail reproduction. This observation
	provides additional justification for recomputing all final candidate
	objectives and reliability statistics using HF simulation.
	
	% NOTE: figs/Fig_Sensitivity.png is referenced by name but does not exist in
	% the current figs/ directory (confirmed by directory listing). The
	% \includegraphics call is commented out so this document stays compilable;
	% supply the real file and uncomment it.
%	\begin{figure}
%		\centering
%		%\includegraphics[width=0.85\linewidth]{figs/Fig_Sensitivity.png}
%		\fbox{\parbox{0.8\linewidth}{\centering\vspace{2.2cm}Figure pending: \texttt{Fig\_Sensitivity.png} not found in \texttt{figs/}\vspace{2.2cm}}}
%		\caption{Maximum absolute change in expected loss across all tested
%			perturbations, relative to the TOPSIS-selected design. PCC placement
%			dominates every other variable (Section~\ref{sec:results_sensitivity}).}
%		\label{fig:current_sensitivity}
%	\end{figure}

		\begin{figure}
		\centering
		\includegraphics[width=0.9\linewidth,height=6.0cm]{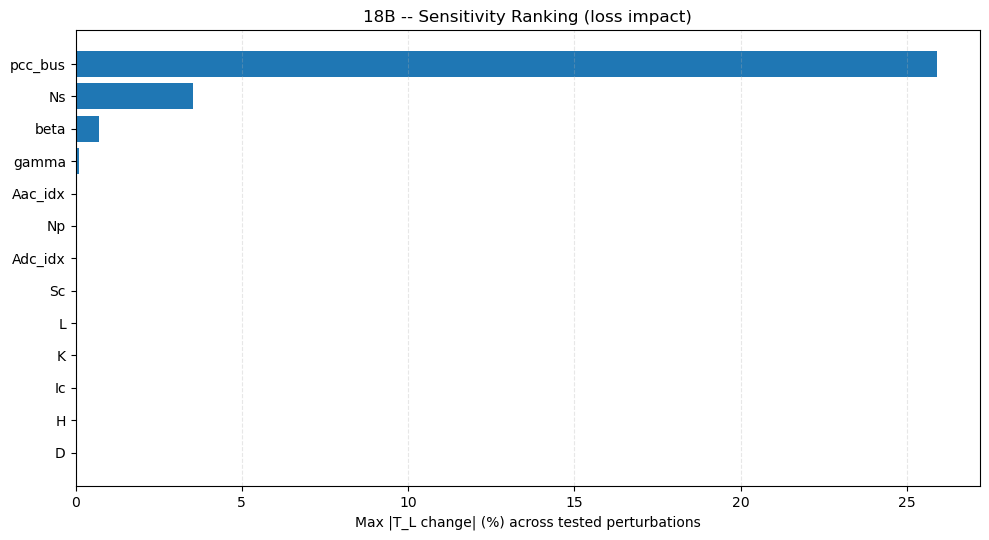}
		\caption{Maximum absolute change in expected loss across all tested
			perturbations, relative to the TOPSIS-selected design. PCC placement
			dominates every other variable (Section~\ref{sec:results_sensitivity}).}
		\label{fig:current_sensitivity}
		\end{figure}
	
	\begin{figure}
		\centering
		\includegraphics[width=0.9\linewidth,height=4.0cm]{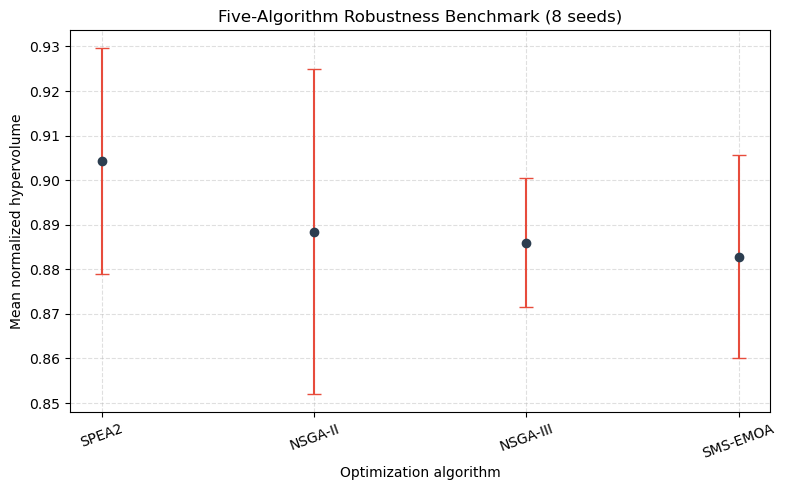}
		\caption{Mean normalized hypervolume $\pm$ one standard deviation, four
			successfully-paired algorithms. Differences are not statistically
			significant (Friedman $p=0.392$; Section~\ref{sec:results_algorithms}).}
		\label{fig:current_algo_hv}
	\end{figure}

	%=====================================================================
	\setcounter{equation}{0}
	\setcounter{table}{0}
	\setcounter{figure}{0}
	\section{Table of Abbreviations}
	\label{appabbrev}
	%=====================================================================
	
\begin{table*}
	\centering
	\caption{Abbreviations and technical terminology used in this study.}
	\label{tab:abbrev}
	\scriptsize
	\setlength{\tabcolsep}{2.2pt}
	\renewcommand{\arraystretch}{0.92}
	
	\begin{tabular}{@{}lp{0.27\textwidth}lp{0.27\textwidth}@{}}
		\toprule
		\textbf{Term} & \textbf{Definition} &
		\textbf{Term} & \textbf{Definition} \\
		\midrule
		
		PV & Photovoltaic &
		HF & High-fidelity reference simulator \\
		
		AC & Alternating current &
		DC & Direct current \\
		
		PF & Power flow &
		PCC & Point of common coupling \\
		
		pu & Per unit &
		SAA & Sample-average approximation \\
		
		MPP & Maximum power point &
		MPPT & Maximum power point tracking \\
		
		STC & Standard test conditions &
		NOCT & Nominal operating cell temperature \\
		
		I--V & Current--voltage characteristic &
		P--V & Power--voltage characteristic \\
		
		GHI & Global horizontal irradiance &
		DNI & Direct normal irradiance \\
		
		DHI & Diffuse horizontal irradiance &
		POA & Plane-of-array irradiance \\
		
		SNN & State Neural Network &
		SNN-DATA & Data-only State Neural Network \\
		
		PINS & Physics-Informed Neural Surrogate &
		MLP & Multilayer perceptron \\
		
		GELU & Gaussian Error Linear Unit &
		AI & Artificial intelligence \\
		
		NSGA-II & Non-dominated Sorting Genetic Algorithm II &
		NSGA-III & Non-dominated Sorting Genetic Algorithm III \\
		
		SPEA2 & Strength Pareto Evolutionary Algorithm 2 &
		SMS-EMOA & S-Metric Selection Evolutionary Multiobjective Algorithm \\
		
		MOEA/D & Multiobjective Evolutionary Algorithm based on Decomposition &
		MOPSO & Multiobjective Particle Swarm Optimization \\
		
		TOPSIS & Technique for Order Preference by Similarity to Ideal Solution &
		CP95 & One-sided 95\% Clopper--Pearson lower bound \\
		
		FF & False-feasible rate &
		FI & False-infeasible rate \\
		
		MSE & Mean-squared error &
		MAE & Mean-absolute error \\
		
		RMSE & Root-mean-square error &
		NRMSE & Normalized root-mean-square error \\
		
		$R^2$ & Coefficient of determination &
		SD & Standard deviation \\
		
		HV & Hypervolume &
		GD & Generational distance \\
		
		IGD & Inverted generational distance &
		MIDAS & Michigan Institute for Data and AI in Society \\
		
		\bottomrule
	\end{tabular}
\end{table*}

	% Author biographies are intentionally omitted from this initial-submission
	% candidate to help meet TSG's 10-page limit (standard practice: bios carry no
	% technical content and are customarily added only once a paper is accepted
	% for publication). Full four-author biography blocks (education, affiliation,
	% research interests, ~100-150 words each) are retained in the extended
	% version file and should be reinstated here before final typesetting.
	
	\end{document}